\documentclass[12pt]{article}
\usepackage{a4wide}
\usepackage[centertags]{amsmath}
\usepackage{amssymb}
\usepackage[sort&compress,numbers]{natbib}
\usepackage{ifpdf}
\usepackage{setspace}
\usepackage{amsfonts}
\usepackage{color}
\usepackage{threeparttable}
\ifpdf
\usepackage[colorlinks=true,
      linkcolor=black,
      citecolor=black,
      urlcolor=blue,
      filecolor=blue,
      pdfstartview=FitV,
      pdftitle={},
        pdfauthor={},
        pdfsubject={Hydrodynamics},
        pdfkeywords={hydrodynamics, AdS/CFT},
        pdfpagemode=None,
        bookmarksopen=true
      ]{hyperref}

\else
\usepackage[hypertex]{hyperref}
\fi
\usepackage{rotating}
\makeatletter
\@addtoreset{equation}{section}

\makeatletter
\renewcommand\section{\@startsection {section}{1}{\z@}%
                                   {-3.5ex \@plus -1ex \@minus -.2ex}
                                   {2.3ex \@plus.2ex}%
                                   {\normalfont\large\bfseries}}
\renewcommand\subsection{\@startsection{subsection}{2}{\z@}%
                                     {-3.25ex\@plus -1ex \@minus -.2ex}%
                                     {1.5ex \@plus .2ex}%
                                     {\normalfont\bfseries}}

 \newcommand{\T}{{\Theta}}

\newcommand{\CG}{\mathcal{G}}
\newcommand{\CF}{\mathcal{F}}
 
 \newcommand{\CO}{\mathcal{O}}

 \newcommand{\CT}{\mathcal{T}}
 \newcommand{\CN}{\mathcal{N}}
 \newcommand{\CD}{\mathcal{D}}

\newcommand{\bs}{{\bar{\sigma}}}
\newcommand{\bt}{{\bar{\tau}}}

\def\sec#1{\S \;\ref{#1}}

\def\req#1{(\ref{#1})}

\def\AdS#1{AdS$_{#1}$}
\def\R{{\bf R}}
\def\ie{{\it i.e.}}
\def\etc{{\it etc.}}
\def\l{\ell}
\def\cf{{\it cf}}

\def\bz{b^{(0)}}
\def\bo{b^{(1)}}

\def\bs{{\bf s}}
\def\bv{{\bf v}}
\def\bt{{\rm \bf {t}}}
\def\bCV{{\mathfrak V}}
\def\bCS{{\mathfrak S}}
\def\bCT{{\mathfrak T}}
\def\Ss#1{\bs{\mathnormal #1}}
\def\ST#1{\bCS{\mathnormal #1}}
\def\V#1{\bv{\mathnormal #1}}
\def\VT#1{\bCV{\mathnormal #1}}
\def\T#1{\bt{\mathnormal #1}}
\def\TT#1{\bCT{\mathnormal #1}}

\title{{\bf Nonlinear Fluid Dynamics from Gravity}}

\author{Sayantani Bhattacharyya$^a$\footnote{sayanta@theory.tifr.res.in}, \ Veronika E Hubeny$^b$\footnote{veronika.hubeny@durham.ac.uk},  \\[1.5mm]
Shiraz Minwalla$^a$\footnote{minwalla@theory.tifr.res.in}, \ Mukund Rangamani$^b$\footnote{mukund.rangamani@durham.ac.uk} \\ \\
\small{\emph{$^{a}$Department of Theoretical Physics,Tata Institute of Fundamental Research,}} \\
\small{\emph{Homi Bhabha Rd, Mumbai 400005, India}} \\ [1mm]
\small{\emph{$^{b}$  Centre for Particle Theory \& Department of
Mathematical Sciences}}, \\
\small{\emph{Science Laboratories, South Road, Durham DH1 3LE, United Kingdom}}
}

\begin{document}

\setlength{\baselineskip}{16pt}
\begin{titlepage}
\maketitle
\begin{picture}(0,0)(0,0)
\put(350,335){TIFR/TH/07-44}
\put(350, 320){DCPT-07/73}
\put(350,305){NI07097}
\end{picture}
\vspace{-36pt}

\begin{abstract}
Black branes in AdS$_5$ appear in a four parameter family labeled by
their velocity and temperature. Promoting these parameters to Goldstone
 modes or collective coordinate fields -- arbitrary functions of the 
coordinates on the boundary of AdS$_5$ -- we use Einstein's equations together with regularity requirements and boundary conditions to determine their dynamics. The resultant equations turn out to be those of boundary fluid dynamics, with specific values for fluid parameters. Our analysis is perturbative in the boundary derivative 
expansion but is valid for arbitrary amplitudes. Our work may be regarded 
as a derivation of the nonlinear equations of boundary fluid dynamics from 
gravity. As a concrete application we find an explicit expression for the 
expansion of this fluid stress tensor including terms up to second 
order in the derivative expansion.
 \end{abstract}
\thispagestyle{empty}
\setcounter{page}{0}
\end{titlepage}

\renewcommand{\baselinestretch}{1}  
\tableofcontents
\renewcommand{\baselinestretch}{1.2}  
\section{Introduction}
\label{intro}

The AdS/CFT correspondence provides an important laboratory to explore 
both gravitational physics as well as strongly coupled dynamics in a 
class of quantum field theories. Using 
this correspondence it is possible to test general lore about 
quantum field theory in a non perturbative setting and so learn 
general lessons about strongly coupled dynamics. Conversely, it is also 
possible to use the AdS/CFT duality to convert strongly held convictions 
about the behaviour of quantum field theories into general lessons about 
gravitational and stringy dynamics. 

In this paper we use the AdS/CFT correspondence to study the effective 
description of strongly coupled conformal field theories at long 
wavelengths. On physical grounds it is reasonable that any interacting quantum 
field theory equilibrates locally at high enough energy densities,  
and so admits an effective description in terms of fluid 
dynamics. The variables of such a description are the local 
densities of all conserved charges together with the local 
fluid velocities. The equations of fluid dynamics are simply the equations 
of local conservation of the corresponding charge currents, supplemented 
by  constitutive relations that express these currents as functions of fluid 
mechanical variables. As fluid dynamics is a long wavelength effective theory, 
these constitutive relations are usually specified in a derivative expansion. 
At any given order, thermodynamics plus symmetries determine 
the form of this expansion up to a finite number of undetermined 
coefficients. These coefficients may then be obtained either from 
measurements or from microscopic computations.

The best understood examples of the AdS/CFT correspondence relate the 
strongly coupled dynamics of certain conformal field theories to the 
dynamics of gravitational systems in AdS spaces. 
In this paper we will demonstrate that Einstein's equations with a negative 
cosmological constant, supplemented with appropriate regularity 
restrictions and boundary conditions, reduce to the nonlinear equations 
of fluid dynamics in an appropriate regime of parameters. We provide 
a systematic framework to construct this universal nonlinear fluid dynamics, 
order by order in a boundary derivative expansion.  Our work builds on  
earlier derivations of linearized fluid dynamics from linearized gravity 
by Policastro, Son and Starinets \cite{Policastro:2001yc} and on earlier examples of the duality between nonlinear fluid dynamics and gravity by Janik, 
some of the current authors and collaborators \cite{Janik:2005zt,Janik:2006gp, Nakamura:2006ih, Bhattacharyya:2007vs} (\cf, \cite{Sin:2006pv, Janik:2006ft, Friess:2006kw, Kajantie:2006ya, Heller:2007qt,Kajantie:2007bn} for related work and extensions and  \cite{Chesler:2007sv,Benincasa:2007tp} for some recent work). There is a large literature in deriving linearized hydrodynamics from AdS/CFT, see \cite{Herzog:2002fn,Policastro:2002tn,Policastro:2002se,Son:2002sd,Herzog:2002pc,Herzog:2003ke,Kovtun:2003wp,Buchel:2003tz,Buchel:2004di,Buchel:2004qq,Kovtun:2004de,Kovtun:2005ev,Benincasa:2005iv,Maeda:2006by,Mas:2006dy,Saremi:2006ep,Son:2006em,Benincasa:2006fu} for developments in this area and \cite{Son:2007vk} for a review and comprehensive set of references.

Our results, together with those of earlier papers referred to above, 
may be interpreted from several points of view. First, one may view them 
as a confirmation that fluid dynamics is the correct long wavelength 
effective description of strongly coupled field theory dynamics. Second, 
one could assume the correctness of the fluid description and  view our 
results as providing information on the allowed singularities of `legal' solutions
of gravity. Finally, our work may be used  to extract the values of all coefficients of the various terms  in the expansion of the stress tensor in the fluid dynamical derivative  expansion, for the fluid dual to gravity on \AdS{5}.  The 
universal behaviour of the shear viscosity - a coefficient 
of a term in the expansion of the stress tensor to first order in field theory 
derivatives - in fluids dual to gravity \cite{Kovtun:2004de} has already 
attracted attention and has impacted experimental analysis of RHIC data \cite{Shuryak:2003xe,Shuryak:2004cy,Shuryak:2006se}. 
In this paper we work out the universal values of all coefficients of 
(nonlinear) two derivative terms stress tensor of the distinguished 
conformal  fluid dual to gravity on \AdS{5}.

Consider any two derivative theory of  five dimensional gravity 
interacting with other fields, that has \AdS{5} as a solution.
Examples of such theories include IIB supergravity on \AdS{5} $\times M$
where $M$ is any compact five dimensional Einstein manifold with positive cosmological constant; for example $M=S^5$, $T^{1,1}$  and  $Y^{p, q}$ for all $p, q$. The solution space of such systems has a universal sub-sector; the solutions of pure gravity with a negative cosmological constant.\footnote{Recall that the Einstein frame Lagrangian contains no interaction terms  that are linear in the non gravitational fluctuations.} We will focus on this universal sub-sector in a particular long wavelength limit. Specifically, we study all solutions that tubewise\footnote{We will work in AdS spacetimes where the radial coordinate $r \in (0,  \infty)$ and will refer to the remaining coordinates $x^\mu = (v,x_i) \in \R^{1,3}$ as field theory or boundary coordinates. The  tubes referred to in the text cover a small patch in field theory directions, but include all values of $r$ well separated from the black brane singularity at $r=0$; typically $r \ge r_h$ where $r_h$ is  the scale set by the putative horizon.} approximate black branes in \AdS{5}, whose temperature and boost velocities vary as a function of boundary  coordinates $x^\mu$ on a length scale that is large compared to the inverse temperature of the brane. We investigate all such
solutions order by order in a perturbative expansion; the perturbation parameter is  the length scale of boundary variation divided by the thermal length scale. Within the domain of validity of our perturbative procedure (and subject to 
a technical assumption), we establish the existence of a one to one map between these gravitational solutions and the solutions of the equations of a distinguished system of boundary conformal fluid dynamics. Implementing our perturbative procedure to second order, we explicitly construct the fluid dynamical stress tensor of this distinguished fluid to second order in the derivative expansion.

Roughly speaking, our construction may be regarded as  the `Chiral Lagrangian' 
for brane horizons. Recall that the isometry group of \AdS{5} is $SO(4,2)$. The Poincare algebra plus dilatations form a distinguished subalgebra of this group; one that acts mildly on the boundary. The rotations $SO(3)$ and translations $\R^{3,1}$ that belong to this subalgebra annihilate the static black brane solution in \AdS{5}. However the remaining symmetry generators -- dilatations and boosts --
act nontrivially on this brane, generating a 4 parameter set of brane
solutions. These four parameters are simply the temperature and the velocity
of the brane. Our construction effectively promotes these parameters to `Goldstone fields' (or perhaps more accurately collective coordinate fields) and determines the 
effective dynamics of these collective coordinate fields, order by order in the derivative expansion, but making no assumption about amplitudes. Of course the collective coordinates method has a distinguished tradition in theoretic physics; see for instance the derivation of the Nambu-Goto action in \cite{Gervais:1974db}. Our paper, which applies these methods to black brane horizons, is strongly reminiscent of the membrane paradigm of black hole physics, and may perhaps be regarded as the precise version of this paradigm in its natural setting, \ie, AdS spacetime.  

Seen from inverse point of view, our construction may be regarded as a
map from solutions of the relativistic fluid dynamics equations on $\R^{3,1}$
to the space of  long wavelength, locally black brane, solutions of
gravity in \AdS{5}. That is, we present a systematic procedure to explicitly
construct a metric dual to any solution of the equations of the
distinguished fluid dynamics alluded to above.  This metric solves the Einstein's 
equations to a given order in the derivative expansion (one higher than the order to which the equations of fluid dynamics were formulated and solved), asymptotes to \AdS{5} with a boundary stress tensor equal to the fluid dynamical stress tensor, and is regular away  from the usual singularity of black branes (chosen by convention to be at $r=0$).

As an important physical input into our procedure, we follow 
 \cite{Janik:2006gp, Janik:2006ft,Heller:2007qt} to demand that all the solutions we study are  regular away from the $r=0$ curvature singularity of 
black branes, and in particular at the the location of the horizon of 
the black brane tubes out of which our solution is constructed. 
 We present our construction in the analogue of Eddington-Finklestein coordinates which extend all the way to the future curvature 
singularity. Although we have not yet performed a careful 
global analysis of our solutions, it seems rather clear that they each possess 
a  regular event horizon that shields the boundary from this curvature 
singularity. 

This paper is organized as follows. We begin in \sec{preamble} with the basic outline of the computation expanding on the ideas presented above. In \sec{pertexp} we outline in detail the logic and strategy of our perturbative procedure. We then proceed in \sec{metstrf} to  implement our perturbative procedure to first order in the derivative expansion. In \sec{metstrs} we extend our computation to second order in the same expansion. In \sec{secfluid} we demonstrate the Weyl invariance of the fluid dynamical stress tensor we obtain, and further use this stress tensor to compute corrections to the dispersion relation for sound and shear waves in this fluid. In \sec{discuss} we end with a discussion of our results and of future directions.

\paragraph{Note added:} After we had completed writing this paper we learnt of related work soon to appear \cite{Baier:2007fj}. The authors of this paper utilize Weyl invariance to constrain the form of the second order fluid dynamical stress tensor up to 5 undetermined coefficients. They then use information from linearized gravitational quasinormal mode calculations together with an earlier computation of Janik and collaborators to determine 3 of these five coefficients. As far as we have been able to tell, their results are consistent with the full second order stress tensor (and prediction for quasinormal mode frequencies) presented herein. This is a nontrivial check of our results.  We thank the authors of \cite{Baier:2007fj} for sharing their results with us prior to publication.
\\        
{\bf Note added in  v2:} The preprint \cite{Baier:2007fj} appeared
simultaneously in an arXiv listing with ${\bf v1}$ of this paper. In 
\sec{sechydro} of  this updated version of our preprint we have 
presented a detailed comparison of our results with those of 
\cite{Baier:2007fj}; where they overlap we find perfect agreement.

\section{Fluid dynamics from gravity}
\label{preamble}

We begin with a description of  the procedure we use  to construct a map from solutions of fluid dynamics to solutions of gravity. We then summarize the results obtained by implementing this procedure to second order in the derivative expansion.

Consider a theory of pure gravity with a negative cosmological constant. With a particular choice of units  ($R_{AdS} = 1$)  Einstein's equations are
given by\footnote{We use upper case Latin indices $\{M,N, \cdots\}$ to denote bulk directions, while lower case Greek indices $\{\mu ,\nu, \cdots\}$ refer to field theory or boundary directions. Finally, we use lower case Latin indices $\{i,j,\cdots\}$ to denote the spatial directions in the boundary.}
\begin{equation}\label{ein} \begin{split}
&E_{M N}= R_{M N} - \frac{1}{2} g_{M N} R- 6\, g_{MN}=0\\
\implies &R_{M N} + 4\, g_{M N}=0, \qquad R=-20 . 
\end{split}
\end{equation}

Of course the equations \eqref{ein} admit \AdS{5} solutions. Another class of solutions to these equations is given by the `boosted black branes'\footnote{The indices in the boundary  are  raised and lowered
with the Minkowski metric \ie, $u_{\mu} = \eta_{\mu \nu} \, u^\nu$. }
\begin{equation}\label{boostedbrane}
ds^2 =-2\, u_{\mu}\, dx^{\mu} dr -r^2\, f(b\,r)\, u_{\mu}\, u_{\nu}\, dx^{\mu}dx^{\nu} + r^2\, P_{\mu\nu}\, dx^{\mu} dx^{\nu} \ , 
\end{equation}
with
\begin{equation} \label{defun} \begin{split}
f(r) & =1- \frac{1}{ r^4} \\
u^v&=\frac{1}{\sqrt{1-\beta^2}} \\
u^i&=\frac{\beta_i}{\sqrt{1-\beta^2}} \ ,
\end{split}
\end{equation}
where the temperature  $T = {1\over \pi\, b}$ and velocities $\beta_i$ are all constants with $\beta^2 = \beta_j \, \beta^j$, and
\begin{equation}\label{defp}
P^{\mu \nu}= u^\mu u^\nu +\eta^{\mu \nu} 
\end{equation}
is the projector onto spatial directions. The metrics \eqref{boostedbrane} describe the uniform black brane written in ingoing Eddington-Finkelstein coordinates, at temperature $T$, moving at velocity $\beta^i$.\footnote{ As we have explained above, the 4 parameter set of metrics \eqref{boostedbrane} may all be obtained
from 
\begin{equation} \label{blackbrane}
ds^2 = 2\, dv\, dr - r^2 \, f(r) \, dv^2 + r^2 \,d{\bf x}^2 \ , \end{equation}
with $f=1-\frac{ 1 }{r^4}$ via a coordinate transform. The coordinate transformations in question are generated by a subalgebra of the isometry group of 
 \AdS{5}.}

Now consider the metric \eqref{boostedbrane} with the constant parameter $b$
and the velocities $\beta_i$ replaced by slowly varying functions
$b(x^\mu ), \beta_i(x^\mu)$ of the boundary coordinates.
\begin{equation}\label{boostedg0}
ds^2 =-2\,u_{\mu}(x^\alpha)\,dx^{\mu} \,dr -r^2\, f\left(b(x^{\alpha})\,r\right) \,
u_{\mu}(x^\alpha) \, u_{\nu} (x^\alpha) \, dx^{\mu}\, dx^{\nu} +
r^2\, P_{\mu\nu}(x^\alpha) \, dx^{\mu} \, dx^{\nu} \ .
\end{equation}
Generically, such a metric (we will denote it by  $g^{(0)}(b(x^\mu), \beta_i(x^\mu)$)
is not a solution to Einstein's equations. Nevertheless it has two attractive
features. Firstly, away from $r=0$, this deformed metric is everywhere
non-singular. This pleasant feature is tied to our use of
Eddington-Finkelstein\footnote{It is perhaps better to call these generalized Gaussian null coordinates as they are constructed with the aim of having the putative horizon  located at the hypersurface $r(x^\mu) = r_h$.} coordinates.\footnote{A similar ansatz for a black branes in (for instance) Fefferman-Graham  coordinates \ie, Schwarzschild like coordinates respecting Poincar\'e symmetry, is singular at $r\, b=1$.} Secondly, if all derivatives of the parameters $b(x^\mu)$ and $\beta_i(x^\mu)$ are small, $g^{(0)}$ is tubewise\footnote{As explained 
above, any given tube consists of all values of $r$ well separated from $r=0$, but only a small region of  the boundary coordinates $x^\mu$.} well approximated by a boosted black brane. Consequently, for slowly varying functions $b(x^\mu)$,  $\beta_i(x^\mu)$, it might seem intuitively plausible that \eqref{boostedg0} is a good approximation to a true solution of Einstein's equations with a regular event horizon. The main result of our paper is that this intuition is correct, provided the functions $b(x^\mu)$ and $\beta_i(x^\mu)$ obey
a set of equations of motion, which turn out simply to be the equations of
boundary fluid dynamics.

Einstein's equations, when evaluated on the metric $g^{(0)}$, yield terms
of first and second order in field theory (\ie, $(x_i, v) \equiv x^\mu$)
derivatives of the temperature and velocity fields.\footnote{As $g^{(0)}$ is an exact solution to Einstein's equations when these fields are constants, terms
with no derivatives are absent from this expansion.} By performing a
scaling of coordinates to set $b$ to unity (in a local patch), it is possible to show
that field theory derivatives of either $\ln b(x^\mu)$ or $\beta_i(x^\mu)$ always
appear together with a factor of $b$. As a result, the contribution of $n$ derivative 
terms to the Einstein's equations is suppressed (relative to
terms with no derivatives) by a factor of $(b/L)^n \sim 1/(T\, L)^n$. Here
$L$ is the length scale of variations of the temperature and velocity
fields in the neighbourhood of a particular point, and $T$ is the temperature
at that point. Therefore, provided $L\,T \gg 1$, it is sensible to solve Einstein's
equations perturbatively in the number of field theory derivatives.\footnote{Note that the  variation in the radial direction, $r$, is never slow. Although we work order by order in the field theory derivatives, we will always solve all differential equations in the $r$ direction exactly.}

In \sec{pertexp} we formulate the perturbation theory described in the previous paragraph, and explicitly implement this expansion to second order in $1/(L\,T)$. As we have mentioned above
it turns out to be possible to find a gravity solution dual to a boundary
velocity and temperature profile only when these fields obey the equation of
motion
\begin{equation} \label{encon}
\partial_\mu T^{\mu\nu}=0
\end{equation}
where the rescaled\footnote{Throughout this paper $T^{\mu \nu}= 
16 \pi\, G_5 \,t^{\mu \nu}$ where $G_5$ is the five dimensional Newton and 
$t^{\mu \nu}$ is the conventionally defined stress tensor, \ie, the charge 
conjugate to translations of the coordinate $v$.} 
stress tensor $T^{\mu \nu}$ (to second order in derivatives) is given by
\begin{equation} \label{fmst} \begin{split}
T^{\mu\nu}=& (\pi \,T)^4
\left( \eta^{\mu \nu} +4\, u^\mu u^\nu \right) -2\, (\pi\, T )^3 \,\sigma^{\mu \nu} \\
& + (\pi T)^2 \,\left( \left(\ln 2\right) \, T_{2a}^{\mu \nu} +2\, T_{2b}^{\mu\nu} +
\left( 2- \ln2 \right)  \left[ \frac{1}{3} \, T_{2c}^{\mu \nu}
+ T_{2d}^{\mu\nu} + T_{2e}^{\mu\nu} \right] \right) \\
\end{split}
\end{equation}
where
\begin{equation}\label{defst}
\begin{split}
\sigma^{\mu\nu}&= P^{\mu \alpha} P^{\nu \beta} \, 
\, \partial_{(\alpha} u_{\beta)}
-\frac{1}{3} \, P^{\mu \nu} \, \partial_\alpha u^\alpha \\
T_{2a}^{\mu\nu}&= \epsilon^{\alpha \beta \gamma (\mu} \, \sigma_{\;\;\gamma}^{\nu)} \, u_\alpha \, \l_\beta \\
T_{2b}^{\mu\nu}&= \sigma^{\mu\alpha} \sigma_{\;\alpha}^{\nu} - \frac{1}{3}\,P^{\mu\nu}\, \sigma^{\alpha \beta } \sigma_{\alpha \beta} \\
T_{2c}^{\mu\nu}&=\partial_\alpha u^\alpha\,\sigma^{\mu\nu}\\
T_{2d}^{\mu\nu}&=  \CD u^\mu \,   \CD
 u^\nu  - \frac{1}{3}\, P^{\mu\nu}\, \CD u^\alpha  \, \CD u_\alpha  \\
  T_{2e}^{\mu \nu}&= P^{\mu\alpha} \, P^{\nu\beta}\, \CD \left(\partial_{(\alpha} u_{\beta)}\right)
 - \frac{1} {3}  \,P^{\mu \nu}\,P^{\alpha \beta} \, \CD \left(\partial_\alpha u_\beta  \right) \\
 \l_\mu &=\epsilon_{\alpha \beta \gamma \mu} \, u^\alpha \partial^\beta u^\gamma .
 \end{split}
 \end{equation}
 Our conventions are  $\epsilon_{0123} = -\epsilon^{0123} =  1$ and 
$\CD \equiv u^\alpha \partial_\alpha$ and the brackets $()$ around the indices to  denote 
symmetrization, \ie, $a^{(\alpha} b^{\beta)}=
(a^\alpha b^\beta + a^\beta b^\alpha)/2$.  

These constraints are simply the equations of fluid dynamics expanded to
second order in the derivative expansion. The first few terms in the expansion
\eqref{fmst} are familiar. The derivative free terms describe a perfect fluid
with pressure (\ie, negative free energy density) $\pi^4 \,T^4$, and so (via
thermodynamics) entropy density $s=4 \pi^4\, T^3$. The viscosity $\eta$ of this
fluid may be read off from the coefficient of $\sigma^{\mu\nu}$ and is given
by $\pi^3\, T^3$. Notice that $\eta/s=1/(4 \pi)$, in agreement with the
famous result of Policastro, Son and  Starinets \cite{Policastro:2001yc}.

Our computation of the two derivative terms in \eqref{fmst} is new; the 
coefficients of these terms are presumably related to the various `relaxation
times' discussed in the literature (see for instance \cite{Muronga:2003ta}). 
As promised earlier, the fact that we 
are dealing with a particular conformal fluid, one that is dual to 
gravitational dynamics in asymptotically AdS spacetimes, leads to the 
coefficients being determined as fixed numbers. It would be interesting  
to check whether the stress tensor determined above fits into the framework 
of the so called Israel-Stewart formalism \cite{Israel:1979wp}  
(see  \cite{Muronga:2003ta, Andersson:2007gf} for reviews). 
R. Loganayagam \cite{loga} is currently investigating this issue.

In \sec{secweyl} we have checked that the minimal covariantization of the 
stress tensor \eqref{fmst} 
transforms as $T^{\mu \nu} \rightarrow e^{-6 \phi} \, T^{\mu \nu} $ under the 
Weyl 
transformation  $\eta_{\mu \nu} \rightarrow e^{2 \phi} \, \eta_{\mu \nu} $, $T \rightarrow e^{-\phi} \, T $, $u^{\alpha} \rightarrow e^{-\phi}\, u^\alpha$, for an arbitrary function $\phi(x^\mu)$.\footnote{R. Loganayagam \cite{loga} informs us that he has succeeded in rewriting our stress tensor in a number of different compact forms, one of which makes its covariance under Weyl transformations manifest.} This transformation (together with the manifest tracelessness
of $T^{\mu \nu}$) ensures Weyl invariance of the fluid dynamical equation
\eqref{encon}. Note that we have computed the fluid dynamical
stress tensor only in flat space. The generalization of our expression
above to an arbitrary curved space could well include contributions
proportional to the spacetime curvature tensor. The fact that 
\eqref{fmst} is Weyl invariant by itself is a bit of a (pleasant) surprise. 
It implies that that the sum of all curvature dependent contributions 
to the stress tensor must be independently Weyl invariant.

\section{The perturbative expansion}
\label{pertexp}

As we have described in \sec{preamble}, our goal is to set up a
perturbative procedure to solve Einstein's equations in asymptotically AdS spacetimes
order by order in a boundary derivative expansion. In this section we will
explain the structure of this perturbative expansion, and outline
our implementation of  this expansion to second order, leaving the
details of computation to future sections.

\subsection{The basic set up}
\label{setup}

In order to mathematically implement our perturbation theory, 
it is useful to regard $b$ and $\beta_i$ described in \sec{preamble} as functions of the rescaled field theory coordinates $\varepsilon \, x^\mu$ where
$\varepsilon$ is a formal parameter that will eventually be set to unity.
Notice that every derivative of $\beta_i$ or $b$ produces a power of
$\varepsilon$, consequently powers of $\varepsilon$ count the number of derivatives. We now describe a procedure to solve Einstein's equations in
a power series in $\varepsilon$. Consider the metric\footnote
{For convenience of notation we are dropping the spacetime indices in 
$g^{(n)}$. We also suppress the dependence of $b$ and $\beta_i$ on $x^\mu$.}
\begin{equation}\label{perturbansatz}
g = g^{(0)}(\beta_i, b) + \varepsilon \,g^{(1)}(\beta_i, b) + \varepsilon^2 \, g^{(2)}(\beta_i, b) + \CO\left(\varepsilon^3\right) ,
\end{equation}
where $g^{(0)}$ is the metric \eqref{boostedg0} and $g^{(1)}, g^{(2)}$ etc are correction metrics that are yet to be determined.  As we will explain below,
perturbative solutions to the gravitational equations exist only when
the velocity and temperature fields obey certain equations of motion.
These equations are corrected order by order in the $\varepsilon$ expansion; this forces us to correct the velocity and temperature fields themselves, order by order
in this expansion. Consequently we set
\begin{equation} \label{veltempexp}
\beta_i= \beta_i^{(0)}+ \varepsilon \, \beta_i^{(1)}+ \CO\left(\varepsilon^2\right), \qquad b=\bz+ \varepsilon \, \bo + \CO\left(\varepsilon^2\right) , 
\end{equation}
where $\beta_i^{(m)}$ and $b^{(n)}$ are all functions of $\varepsilon \,x^\mu$.

In order to proceed with the calculation, it will be useful to fix a gauge.
We work with the `background field'
gauge
\begin{equation} \label{gaugecond}
g_{rr}= 0 \ , \qquad g_{r \mu }\propto u_\mu \ , \qquad {\rm Tr}\left(
(g^{(0)})^{-1} g^{(n)} \right)=0 \;\;\; \forall \; n  > 0 . 
\end{equation}
Notice that the gauge condition at the point $x^\mu$ is given only once we
know $u_\mu(v, x^i)$. In other words, the choice above amounts to choosing
different gauges for different solutions, and is conceptually similar to the
background field gauge routinely used in effective action computations
for non abelian gauge theories.

\subsection{General structure of perturbation theory}
\label{pertstruc}
Let us imagine that we have solved the perturbation theory to the
$(n-1)^{{\rm th}}$ order, \ie, we have determined $g^{(m)}$ for $m\leq n-1$, and
have determined the functions $\beta_i^{(m)}$ and $b^{(m)}$ for $m \leq n-2$.
Plugging the expansion \eqref{perturbansatz} into Einstein's equations,
and extracting the coefficient of $\varepsilon^n$,  we
obtain an equation of the form
\begin{equation} \label{homoop}
H\left[g^{(0)}(\beta^{(0)}_i, \bz)\right] g^{(n)}(x^\mu ) = s_n  .
\end{equation}
Here $H$ is a linear differential operator of second order
in the variable $r$ alone. As $g^{(n)}$ is already of order $\varepsilon^n$,
and since every boundary derivative appears with an additional power of
$\varepsilon$, $H$ is an ultralocal operator in the field theory directions.
It is important to note that $H$ is a differential operator only in the variable $r$ and  does not depend  on the variables $x^\mu$. Moreover, the precise form of this operator at the point $x^\mu$ depends only on the values of $\beta^{(0)}_i$ and $\bz$ at $x^\mu$ but not on the derivatives of these functions at that point. Furthermore,  the operator $H$ is independent of $n$; we have the same homogeneous operator at every order in perturbation theory.

The source term $s_n$  however is different at different orders in perturbation
theory. It is a local expression of $n^{{\rm th}}$ order in boundary derivatives
of $\beta^{(0)}_i$ and $\bz$, as well as of $(n-k)^{{\rm th}}$ order in $\beta_i^{(k)}$, $b^{(k)}$ for all $k \leq n-1$. Note that $\beta_i^{(n)}$ and $b^{(n)}$ do not enter the $n^{{\rm th}}$ order equations as constant (derivative free) shifts of velocities
and temperatures solve the Einstein's equations.

The expressions \eqref{homoop} form a set of $5 \times 6/2=15$ equations. It turns out that four
of these equations do not involve the unknown function $g^{(n)}$ at all; they
simply constrain the velocity functions $b$ and $\beta_i$. There is one 
redundancy among the remaining 11 equations which leaves  10 independent 
`dynamical' equations. These  may be used to solve for the 10 unknown 
functions in our gauge fixed metric correction $g^{(n)}$, as we describe in 
more detail below.  

\subsubsection{Constraint equations}

By abuse of nomenclature, we will refer to those of the Einstein's equations that
are of first order in $r$ derivatives as constraint equations. Constraint
equations are obtained by dotting the tensor $E_{MN}$ with the vector
dual to the one-form $dr$. Four of the five constraint equations (\ie, those whose
free index is a $\mu$ index) have an especially simple boundary
interpretation; they are simply the equations of boundary energy momentum 
conservation. In the context of our perturbative analysis,
these equations simply reduce to
\begin{equation} \label{stcons}  \partial_\mu T_{(n-1)}^ {\;\mu \nu}=0
\end{equation}
where $T_{(n-1)}^ {\;\mu \nu} $ is the boundary stress tensor dual the solution
expanded up to $\CO\left(\varepsilon^{n-1}\right)$. Recall that each of $g^{(0)}, g^{(1)} ... $
are local functions of $b, \beta_i$. It follows that the stress tensor
$T_{(n-1)}^ {\;\mu \nu}$ is also a local function (with at most $n-1$ derivatives)
of these temperature and velocity fields. Of course the stress tensor
$T_{(n-1)}^ {\;\mu \nu}$ also respects 4 dimensional conformal invariance.
Consequently it is a `fluid dynamical' stress tensor with $n-1$ derivatives,
the term simply being used for the most general stress tensor (with $n-1$
derivatives), written as a function of $u^\mu$ and $T$, that respects 
all boundary symmetries.

Consequently, in order to solve the constraint equations at $n^{{\rm th}}$ order 
one must solve the equations of fluid dynamics to $(n-1)^{{\rm th}}$ order. As 
we have already been handed a solution to fluid dynamics at order $n-2$, all 
we need to do is to correct this solution to one higher order. Though the 
question of how one goes about improving this solution is not the topic of
our paper (we wish only to establish a map between the solutions of fluid 
mechanics and gravity, not to investigate how to find the set of all such 
solutions) a few words in this connection may be in order.  The only quantity 
in \eqref{stcons} that is not already known from the results
of perturbation theory at lower orders are $\beta_i^{(n-1)}$ and $b^{(n-1)}$.
The four equations \eqref{stcons} are linear differential equations in these
unknowns that presumably always have a solution. There is a 
non-uniqueness in these solutions given by the zero modes obtained by 
linearizing the equations of stress energy conservation at zeroth order. 
These zero modes may always be absorbed into a redefinition of 
$\beta^{(0)}_i, \bz$, and so do not correspond to a physical non-uniqueness 
(\ie, this ambiguity goes away once you specify more clearly what your 
zeroth order solution really is).  

Our discussion so far may be summarized as
follows: the first step in solving Einstein's equations at $n^{{\rm th}}$ order
is to solve the constraint equations -- this amounts to solving the
equations of fluid dynamics at $(n-1)^{{\rm th}}$ order \eqref{stcons}. As we explain below,
while it is of course difficult in general to solve these differential equations throughout $\R^{3,1}$, it is easy to solve them locally 
in a derivative expansion about any point; this is in fact sufficient 
to implement our ultralocal perturbative procedure.

\subsubsection{Dynamical equations} 

The remaining constraint $E_{rr}$ and the `dynamical' Einstein's equations 
$E_{\mu\nu}$ may be used to solve for the unknown function $g^{(n)}$. Roughly 
speaking, it turns out to  be possible to make a  judicious choice of  variables such that the operator $H$ is converted into 
a decoupled system of first order differential operators.  It is then 
simple to solve the equation \eqref{homoop} for an arbitrary source $s_n$ 
by direct integration. This procedure actually yields a whole linear 
space of solutions. The undetermined constants of integration in this 
procedure are arbitrary functions of $x^\mu$ and multiply  zero modes 
of the operator \eqref{homoop}. As we will see below, for an arbitrary 
non-singular and appropriately normalizable source $s_n$ (of the sort that 
one expects to be generated in perturbation theory\footnote{Provided the solution
at order $n-1$ is non-singular at all nonzero $r$, it is guaranteed to produce a 
non-singular source at all nonzero $r$. Consequently, the non-singularlity of 
$s_n$ follows inductively. We think is possible to make a similar inductive 
argument for the large $r$ behaviour of the source, but have not yet 
formulated this argument precisely enough to call it a proof. }), it is always 
possible to choose these constants to ensure that $g^{(n)}$ is appropriately normalizable 
at $r=\infty$ 
and non-singular at all nonzero $r$. These requirements do not yet completely 
specify the solution for $g^{(n)}$,
as $H$ possesses a set of zero modes that satisfy both these requirements.
A basis for the linear space of zero modes, denoted $g_b$ and $g_i$, 
is obtained by differentiating the 4 parameter class of solutions \eqref{boostedbrane}
with respect to the parameters $b$ and $\beta_i$. In other words these
zero modes correspond exactly to infinitesimal shifts of
$\beta^{(0)}_i$ and $\bz$  and so may be absorbed into a redefinition of these
quantities. They reflect only an ambiguity of convention, and may be
fixed by a `renormalization' prescription, as we will do below.

\subsubsection{Summary of the perturbation analysis}
In summary,  it is always possible to find a physically unique solution
for the metric $g^{(n)}$, which, in turn, yields the form of the $n^{{\rm th}}$
order fluid dynamical stress tensor (using the usual AdS/CFT dictionary).
This process, being iterative, can be used to recover the fluid dynamics stress tensor to any desired order in the derivative expansion.

In \sec{outfirst} and \sec{outsecond}  we will provide a few more details 
of our perturbative procedure, in the context of implementing this procedure to
first and second order in the derivative expansion.

\subsection{Outline of the first order computation}
\label{outfirst}

We now present the strategy to implement the general procedure discussed above to 
first order in the derivative expansion. 

\subsubsection{Solving the constraint equations}
\label{constrsol}

The Einstein constraint equations at first order require that the
zero order velocity and temperature fields obey the equations of perfect
fluid dynamics
\begin{equation}\label{conszeroeth}
\partial_\mu T_{(0)}^{\mu \nu}=0 \ ,
\end{equation}
where up to an overall constant
\begin{equation}\label{conszerost}
T_{(0)}^{\mu \nu}=\frac{1}{(\bz)^4}
\left( \eta^{\mu \nu} + 4 \, u_{(0)}^\mu \, u_{(0)}^\nu \right) .
\end{equation}

While it is  difficult to find the general solution to these equations 
at all $x^\mu$, in order to carry out our ultralocal perturbative procedure 
at a given point $y^\mu$, we only need to solve these constraints to first order in 
a Taylor expansion of the fields $b$ and $\beta_i$ about the point $y^\mu$. 
This is, of course,  easily achieved. The four equations \eqref{conszeroeth} 
may be used to solve for the 4 derivatives of the temperature field at $y^\mu$ 
in terms of first derivatives of  the velocity fields at the same point. 
This determines the Taylor expansion of $b$ to first order about $y^\mu$
in terms of the expansion, to first order, of the field $\beta_i$ about the 
same point. We will only require the first order terms in the Taylor expansion
of velocity and temperature fields in order to compute $g^{(1)}(y^\mu) $. 

\subsubsection{Solving the dynamical equations}

As described in the previous section, we expand Einstein's equations to first 
order and find the equations \eqref{homoop}. Using the `solution' 
of \sec{constrsol}, all source terms may be regarded as functions 
of first derivatives of velocity fields only. The equations \eqref{homoop}
are then easily integrated subject to boundary conditions and we find
\eqref{homoop} is given by
\begin{equation}\label{firstordersoln}
g^{(1)}=g^{(1)}_P+f_b(x_i, v) \, g_b+ f_i(x_j, v) \, g_i , 
\end{equation}
where $g^{(1)}_P$ is a particular solution to \eqref{homoop}, and
 $f_b$ and $f_i$ are a basis for the zero modes of $H$ that  were described 
in the \sec{pertstruc}. Plugging in this solution, the full metric $g^{(0)}+g^{(1)}$, when expanded to order first order in $\varepsilon$, is
\eqref{perturbansatz}
\begin{equation}\label{genmet}
g=g^{(0)} +
\varepsilon \left( g^{(1)}_P+ ( f_b + \bo)  g_b+ (f_i + \beta^{(1)}) g_i \right) ,
\end{equation}
where the four functions of $x^\mu$, $f_b +\bo, f_i + \beta^{(1)}_i$ are all
completely unconstrained by the equations at order $\varepsilon$.

\subsubsection{The `Landau' Frame}

Our solution \eqref{firstordersoln} for the first order metric has a four 
function non-uniqueness in it. As $f_b$ and $f_i$ may be absorbed into 
$\bo$ and $\beta^{(1)}_i$ this non-uniqueness simply represents an ambiguity of 
convention, and may be fixed by a `renormalization' choice. We describe our 
choice below.

Given $g^{(1)}$, it is straightforward to use the AdS/CFT correspondence 
to recover the stress tensor.  To first order in $\varepsilon$ the 
boundary stress tensor dual to the metric \eqref{genmet} evaluates to
\begin{equation}\label{firstordbst}
T^{\mu\nu}= {1 \over b^4} \left( \eta^{\mu \nu} +4\, u^\mu u^\nu \right)
-{2 \over b^3} T_{(1)}^{\mu \nu}  ,
\end{equation}
where
\begin{equation}\begin{split}
b&=\bz+\varepsilon( \bo+ f_b) \\
\beta_i&= \varepsilon( \beta^{(1)}_i+ f_i)\\
\end{split}
\end{equation}
where $T_{(1)}^{\mu\nu}$, defined by \eqref{firstordbst}, is an expression 
linear in $x^\mu$ derivatives of the velocity fields and temperature fields. 
Notice that our definition of $T_{(1)}^{\mu\nu}$, via \eqref{firstordbst}, 
depends explicitly on the value of the coefficients $f_i, f_b$ of the
homogeneous modes of the differential equation \eqref{homoop}. These
coefficients depend on the specific choice of the particular solution 
$g^{(1)}_P$, which is of course ambiguous up to addition of homogenous 
solutions. Any given solution \eqref{firstordersoln} may be broken up in
many different ways into particular and homogeneous solutions, resulting
in an ambiguity of shifts of the coefficients of  $f_b, f_i$ and thereby an
ambiguity in $T_{(1)}^{\mu \nu}$.
It is always possible to use the freedom
provided by this ambiguity to set $u_{(0) \mu} \,T_{(1)}^{\mu \nu}=0$. This choice
completely fixes the particular solution $g^{(1)}_P$. We adopt  this 
convention for the particular solution and  then simply simply set $g^{(1)}=g^{(1)}_P$ 
\ie, choose $f_b=f_i=0$. $T_{(1)}^{\mu \nu}$ is now  unambiguously defined
and may be evaluated by explicit computation;  it turns out that 
$$ T_{(1)}^{\,\mu \nu}= \sigma^{\mu \nu}.$$

The discussion of the previous paragraph has a natural
generalization to perturbation theory at any order. As the operator $H$ 
is the same at every order in perturbation theory, the ambiguity for the 
solution of $g^{(n)}$ in perturbation theory is always of the form described 
in \eqref{genmet}. We will always fix the ambiguity in this solution by 
choosing $u_{\mu }\, T_{(k)}^{\mu \nu}=0$. The convention dependence of 
this procedure has a well known counterpart in fluid
dynamics; it is simply the ambiguity of the stress tensor under 
field redefinitions of the temperature and $u^\mu$. Indeed this field 
redefinition ambiguity is standardly fixed by precisely the `gauge' 
choice $u_\mu T_{(1)}^{\mu \nu}=0$. This is the so called `Landau frame' 
widely used in studies of fluid dynamics.\footnote{Conventionally, one writes in fluid mechanics the stress tensor as the perfect fluid part and a dissipative part \ie, 
$T^{\mu \nu } = T_{perfect}^{\mu \nu } + T_{dissipative}^{\mu \nu }$. The Landau gauge condition we choose at every order simply amounts to $u_\mu \, T_{dissipative}^{\mu \nu} = 0$.} 

We present the details of the first order computation in \sec{metstrf}  below.

\subsection{Outline of the second order computation}
\label{outsecond}
Assuming that we have implemented the first order calculation described in 
\sec{outfirst}, it is then possible to find a solution to Einstein's equations at the 
next order. In this case care should be taken in implementing the constraints as we 
discuss below. 

\subsubsection{The constraints at second order}

The general discussion of \sec{pertstruc} allows us to obtain the 
second order solution to Einstein's equations once we have solved the 
first order system as outlined in \sec{outfirst}. However, we need to 
confront an important issue before proceeding, owing to the way we have set 
up the perturbation expansion. Of course perturbation theory at second order 
is well defined only once the first order equations have been solved. 
While in principle we should solve these equations everywhere in $\R^{3,1}$,
in the previous subsection we did not quite achieve that; we were content 
to solve the constraint equation \eqref{conszeroeth} only to first order 
in the Taylor expansion about our special point $y^\mu$. While that was 
good enough to obtain $g^{(1)}$, in order to carry out the second order calculation 
we first need to do better; we must ensure that the first order constraint 
is obeyed to second order in the Taylor expansion of the fields $b^{(0)}$ and 
$\beta_i^{(0)}$ about $y^\mu$. That is, we require
\begin{equation} \label{quad}
\partial_\lambda \partial_\mu T_{(0)}^{\mu \nu}\left(y^\alpha\right)=0 .
\end{equation}

Essentially, we require that $T_{(0)}^{\mu \nu}$ satisfy the conservation equation \eqref{conszeroeth} to order $\varepsilon^2$ before we attempt to find the second order stress tensor. In general, we would have need \eqref{conszeroeth} to be satisfied globally before proceeding; however, the ultralocality manifest in our set-up implies that it suffices that the conservation holds only to the order we are working. If we were interested in say the $n^{{\rm th}}$  order stress tensor $T_{(n)}^{\mu\nu}$ we would need to ensure that the  stress tensor up to order $n-1$ satisfies the conservation equation to order $\CO\left(\varepsilon^{n-1}\right)$.

The equations  \eqref{quad} may be thought of as a set of 16 linear constraints on the 
coefficients of the (40+78) two derivative terms involving $\bz$ and 
$\beta^{(0)}_i$. We use these equations 
to solve for 16 coefficients, and treat the remaining coefficients as 
independent. This process is the conceptual analogue of our zeroth order
`solution' of fluid dynamics at the point $y^\mu$  (described in the previous
subsection), obtained by solving for the first derivatives of temperature
in terms  of the first derivatives of velocities. Indeed it is an extension of
that procedure to the next order in derivatives. See \sec{metstrs} for the 
details of the implementation of this procedure. 
In summary, before we even 
start trying to solve for $g^{(2)}$, we need to plug  
a solution of \eqref{quad} into $g^{(0)}+g^{(1)}$ expanded in a Taylor series expansion
about $y^\mu$. Otherwise we would be expanding the second order equations about 
a background that does not solve the first order fluid dynamics.

\subsubsection{Nature of source terms}

As we have explained above, the Einstein's equations, to second order, take the
schematic form described in \eqref{homoop}
\begin{equation} \label{tendord}
H\left[g^{(0)}(\beta^{(0)}_i, \bz)\right] \,g^{(2)}= s_a + s_b
\end{equation}
We have broken up the source term above into two
pieces, $s_a$ and $s_b$, for conceptual convenience. $s_a$ is a local
functional of $\beta^{(0)}_i$ and $\bz$ of up to  second order in field theory
derivatives. Terms contributing to $s_a$ have their origin both in two field
theory derivatives acting on the metric $g^{(0)}$ and exactly one field theory
derivative acting on $g^{(1)}$ (recall that $g^{(1)}$ itself is a local function
of $\beta^{(0)}_i$ and $\bz$ of first order in derivatives).
The source term $s_b$ is new: it arises from first order derivatives of the
velocity and temperature corrections $\beta^{(1)}_i$ and $\bo$.  This has no
analogue in the first order computation.

As we have explained above, $\beta^{(0)}_i, \bz$ are absolutely any functions that obey the equations \eqref{conszeroeth}. In particular, if it turns of that the functions
$\beta^{(0)}_i+ \varepsilon \,\beta^{(1)}_i$ and $\bz+\varepsilon \, \bo$  obey that equation (to first order in $\varepsilon$) then $\beta^{(1)}_i$ and $\bo$ may each simply be set to zero by an appropriate redefinition of $\beta^{(0)}_i$ and $ \bz_i$. This results in a `gauge' ambiguity of the functions $\beta^{(1)}_i, \bo$. In our ultralocal perturbative procedure, we choose to fix this ambiguity by setting
$\beta^{(1)}_i$ to zero (at our distinguished point $y^\mu$)  while leaving
$\bo$ arbitrary.\footnote{ The functions $\bo_i, \beta^{(1)}_i$ have sixteen
independent first derivatives, all but four of which may be fixed by the
gauge freedom. We choose use this freedom to set all velocity derivatives to
zero. }

\subsubsection{Solution of the constraint equations}

With the source terms in place, the procedure to solve for $g^{(2)}$
proceeds in direct imitation of the first order calculation. 
The constraint equations reduce to the expansion to order $\varepsilon$ 
of the equation of conservation of the stress tensor
\begin{equation}\label{eosec}
T^{\mu\nu} = \frac{1}{b^4} \,\left( 4 \,u^\mu u^\nu + \eta^{\mu \nu} \right)
-2\, \varepsilon \,\frac{1}{b^3} \,\sigma^{\mu \nu}
\end{equation}
with $\beta^i=\beta^{(0)}$, $b= \bz+\varepsilon\, \bo$. 
These four equations may be used to solve for the four derivatives
$\partial_\mu \bo$ at $x^\mu$. Consequently the constraint equations plus 
our choice of gauge, uniquely determined the first order
correction of the temperature field $b^{(1)}$ and velocity field $\beta^{(1)}_i$ as a function of the zeroth  order solution. 

Note that the gauge $\beta_i^{(1)}(y^\mu)=0$ may be consistently chosen at 
any one point $y^\mu$, but not at all $x^\mu$. Nonetheless the results 
for $g^{(2)}$ that we obtain using this gauge will, when appropriately 
covariantized be simultaneously applicable to every spacetime point 
$x^\mu$. The reason for this is that all source terms depend on $b^{(1)}$ and 
$\beta_i^{(1)}$ only through the expansion to order $\varepsilon$ of 
$\partial_\mu T^{\mu \nu}=0 $ with $T^{\mu \nu}$ given by \eqref{eosec}. 
Note that this source term is `gauge invariant' (recall that `gauge' 
transformations are simply shifts of $b^{(1)}$ and $\beta_i^{(1)}$ by zero modes 
of this equation). It follows that $g^{(2)}$ determined via this procedure 
does not depend on our choice of gauge, which was made purely for convenience.

\subsubsection{Solving for $g^{(2)}$ and the second order stress tensor}

Now plugging this solution for $\bo$ into the source terms it is
straightforward to integrate \eqref{tendord} to obtain $g^{(2)}$. We fix the
ambiguity in the choice of homogeneous mode in this solution as before, by requiring 
$T_{(2)}^{\mu \nu} \,u_{(0)\nu} =0$. This condition yields a unique
solution for $g^{(2)}$ as well as for the second order correction to the fluid
dynamical stress tensor $T_{(2)}^{\mu \nu}$, giving rise to the result \eqref{defst}.
We present the details of the second order computation in \sec{metstrs}.

In the rest of this paper we will present our implementation of our 
perturbative procedure described above, to first and second order in the 
derivative expansion.  

\section{The metric and stress tensor at first order}
\label{metstrf}

In this section we will determine the solution, to first order in the
derivative expansion. As we have described in \sec{pertexp}, the equations that determine
$g^{(1)}$ at $x^\mu$ are ultralocal; consequently we are able to solve the
problem point by point. It is always possible to choose coordinates
to set $u^\mu=(1,0, 0, 0)$ and $\bz=1$ at any given point $x^\mu$. Making
that choice, the metric \eqref{boostedg0} expanded to first order in derivatives in the neighbourhood of $x^\mu$ (chosen to be the origin of $\R^{3,1}$ for notational simplicity) is given
by
\begin{equation} \label{fote}
\begin{split}
ds_{(0)}^2 &= 2\, dv \, dr -r^2\, f(r)\, dv^2 + r^2\, dx_i \, dx^i \\
&- 2\, x^{\mu}\, \partial_{\mu}\beta^{(0)}_i \, dx^i \, dr -
2\, x^\mu \partial_\mu
\beta^{(0)}_i \, r^2\, (1-f(r))\, dx^i \, dv  - 4 \,\frac{ x^\mu \, \partial_\mu \bz}{r^2}\, dv^2 \ .
\end{split}
\end{equation}
In order to implement the perturbation programme described in the previous
section, we need to find the first order metric $g^{(1)}$ which, when added
to \eqref{fote}, gives a solution to Einstein's equations to first order in
derivatives.

The metric \eqref{fote} together with $g^{(1)}$ has a background piece (the first line in \eqref{fote}) which is simply the metric of a uniform black brane. In addition it has small first derivative corrections, some of which are known (the second line
of \eqref{fote}), and the remainder of which ($g^{(1)}$) we have to determine.
Now note that the background black brane metric preserves a spatial
$SO(3)$ rotational symmetry. This symmetry allows us to
solve separately for the $SO(3)$ scalars, the $SO(3)$ vector and
$SO(3)$  symmetric traceless two tensor (${\bf  5}$) components of $g^{(1)}$ and lies at the heart of the separability of the matrix valued 
linear operator $H$ into a set of ordinary linear operators.

In the following we will discuss each of these sectors separately and determine $g^{(1)}$. Subsequently, in \sec{fglobal} we present the full solution to order $\varepsilon$ and proceed to calculate the stress tensor in \sec{fstress}.

\subsection{Scalars of $SO(3)$}
\label{fscalar}

The scalar components of $g^{(1)}$ are parameterized by the functions
$h_1(r)$ and $k_1(r)$ according to\footnote{In the spatial $\R^3 \subset \R^{3,1}$
we will often for ease of notation, avoid the use of covariant and contravariant indices and adopt a summation convention for repeated indices \ie, $g^{(1)}_{ii} = \sum_{i=1}^3 \, g^{(1)}_{ii}$.} 
\begin{equation} \label{scalarefperttwo}
\begin{split}
g^{(1)}_{ii} (r) &= 3\, r^2 \,h_1(r)\\
g^{(1)}_{vv}(r) &= \frac{k_1(r)}{r^2}\\
g^{(1)}_{vr}(r)&=-\frac{3}{2}\, h_1(r).
\end{split}
\end{equation}
Here $g^{(1)}_{ii}$ and $g^{(1)}_{vr}$ are related to each other by our gauge choice
$Tr((g^{(0)})^{-1} g^{(1)})= 0$.

The scalar Einstein's equations (\ie, those equations that transform as a
scalar of $SO(3)$) may be divided up into constraints and dynamical equations. The constraint equations are obtained by contracting Einstein's equations (the
first line of \eqref{ein}) with the vector dual to the one form $dr$. The first
scalar constraint is
\begin{equation}\label{scalarconstraint}
r^2\,  f(r) \, E_{vr} + E_{vv}=0 \ ,
\end{equation}
which evaluates to
\begin{equation} \label{scaefcons}
\partial_v \bz=  {\partial_i \beta^{(0)}_i \over 3} \ .
\end{equation}
Below, we will interpret \eqref{scaefcons} as the expansion of the fluid dynamical stress energy conservation, expanded to first order.
The second constraint equation,
\begin{equation} \label{scaconstCtwo}
r^2 \, f(r) \,{E}_{rr} + {E}_{vr}=0 \ ,
\end{equation}
leads to
\begin{equation}\label{scaconsDS}
  12 \,r^3 \,h_1(r) + (3r^4-1)\, h_1'(r)
-k_1'(r)= -6\, r^2\, {\partial_i \beta^{(0)}_i \over 3} \ .
\end{equation}

To this set of constraints we need add only one dynamical scalar equation,\footnote{We have explicitly checked that the equations listed here imply that  the second dynamical equation is automatically satisfied.} the simplest of which turns out to be
\begin{equation} \label{scandyn}
5 \, h_1'(r) + r \, h_1''(r)=0 \ .
\end{equation}

The LHS of \eqref{scandyn} and \eqref{scaconsDS} are the restriction of the 
operator $H$ of \eqref{homoop} to the scalar sector. The RHS 
of the same equations are the scalar parts of the source terms $s_1$. Notice 
that $H$ is a first order operator in the variables $h_1'(r)$ and $k_1(r)$. 
Consequently the equation \eqref{scandyn} may be integrated for an 
arbitrary source term. The resulting solution is regular at all nonzero 
$r$ provided that the source shares this property, and the growth $h_1(r)$ 
at infinity is slower than a constant -- the behaviour of a non normalizable 
operator deformation -- provided the source in \eqref{scandyn} grows slower 
than $1/r$ at large $r$. Once $h_1(r)$ has been obtained
$k_1(r)$ may be determined from \eqref{scaconsDS} by integration, for an 
arbitrary source term. Once again, the solution will be 
regular and grows no faster than $r^3$ at large $r$, provided the source 
in that equation is regular and normalizable. The two source terms of this 
subsection satisfy these regularity and growth requirements, and it seems 
clear that this result will extend to arbitrary order in perturbation 
theory (see the next section).

The general solution to the system \eqref{scaconsDS} and \eqref{scandyn}, 
obtained by the integration described above, is 
\begin{equation} \label{scalarsolnS}
h_1(r)=s + \frac{t}{r^4}, \qquad
k_1(r)={2\, r^3 \,\partial_i \beta^{(0)}_i \over 3} + 3 \, r^4 \, s -{ t \over r^4} + u \ ,
\end{equation}
where $s, t$ and $u$ are arbitrary constants (in the variable $r$).
In the solution above, the parameter $s$ multiplies a non
normalizable mode (which represents a deformation of the field theory
metric) and so is forced to zero by our boundary conditions. A linear
combination of the pieces multiplied by $t$ and $u$ is generated by
the action of the coordinate transformation $r'=r\, (1+a/r^4)$ and so
is pure gauge, and may be set to zero without loss of generality. The
remaining coefficient $u$ corresponds to an infinitesimal temperature variation, and is forced to be zero by our renormalization condition on the stress tensor $u_{(0)}^\mu \,T_{\mu \nu}=0$ (see the subsection
on the stress tensor below).  In summary, each of $s, t, u$ may be set
to zero and the scalar part of the metric $g^{(1)}$,  denoted $g^{(1)}_S$, is
\begin{equation}\label{scalargonetwo}
\left(g^{(1)}_S\right)_{\alpha \beta}\, dx^\alpha dx^\beta = \frac{2}{3}\,r \,\partial_i \beta^{(0)}_i \, dv^2 .
\end{equation}

Two comments about this solution are in order. First note that $k_1(r)$ is
manifestly regular at the unperturbed `horizon' $r=1$, as we require.
Second, it grows at large $r$ like $r^3$. This is intermediate
between the $r^0$ growth of finite energy fluctuations and the $r^4$ growth
of a field theory metric deformation. As $g^{(0)} +g^{(1)}$ obeys the Einstein's 
equations to leading order in derivatives, the usual Fefferman-Graham 
expansion assures us that the sum of first order fluctuations in $g^{(0)} +g^{(1)}$
must (in the appropriate coordinate system) die off like $1/r^4$ compared
to terms that appear in the zeroth order metric (this would correspond to
$k_1(r)$ constant at infinity). Consequently the unusually slow fall off
at infinity of our metric $g^{(1)}$ must be compensated for by an equal but
opposite effect from a first order fluctuation piece in the second line
of \eqref{fote}. This indeed turns out to be the case. While an explicit computation of the boundary stress tensor dual to \eqref{fote} yields a result that diverges like $r^3$, this divergence is precisely cancelled when we add $g^{(1)}$ above to the metric, and the correct value of the stress dual to $g^{(0)}+g^{(1)}$ is in fact zero in the scalar
sector, in agreement with our renormalization condition $u_{(0)\mu}\, T^{\mu\nu}=0$.

\subsection{Vectors of $SO(3)$}
\label{fvector}

In the vector channel the relevant Einstein's equations are the constraint
$r^2\, f(r)\, E_{ri}+ E_{vi}=0$ and a dynamical equation which can be chosen to
be any linear combination of the Einstein's equations $E_{ri}=0$ and $E_{vi}=0$.  The constraint evaluates to
\begin{equation}\label{momcons}
\partial_i \bz=\partial_v \beta^{(0)}_i \ ,
\end{equation}
which we will later interpret as a consequence of the conservation of
boundary momentum.
In order to explore the content of the dynamical equation (we
choose  $E_{ri}=0$),  it is
convenient to parameterize the vector part of the fluctuation metric by
the functions $j^{(1)}_i$, as
\begin{equation}\label{metchangethreetwo}
\left(g^{(1)}_V\right)_{\alpha \beta} dx^\alpha dx^\beta= 2\,r^2\, \left(1 - f(r)\right) \, j^{(1)}_i(r)\,  dv \, dx^i.
\end{equation}

The dynamical equation for $j_i(r)$ turns out to be
\begin{equation}\label{vecdyn}
\frac{d}{dr}\left({ 1\over r^3} \, {d\over dr} j^{(1)}_i(r)\right) = -\frac{3}{r^2} \; \partial_v\beta^{(0)}_i  .
\end{equation}
The LHS of \eqref{vecdyn} is the restriction of the 
operator $H$ of \eqref{homoop} to the vector sector, and the RHS of this 
equation is the projection of $s_1$ to the vector sector. $H$ is of first 
order in the variable $j^{(1)'}(r)$ and so may be integrated for an arbitrary 
source term. The resulting solution is regular and normalizable provided 
the source is regular and decays at infinity faster than $1/r$. This 
condition is obeyed in \eqref{vecdyn}; it seems rather clear that it will 
continue to be obeyed at arbitrary order in perturbation theory (see the 
next section).

Returning to \eqref{vecdyn}, the general solution of this equation is 
\begin{equation} \label{soldyn}
j^{(1)}_i(r)= \partial_v\beta^{(0)}_i\, r^3 + a_i\,  r^4 + c_i
\end{equation}
for arbitrary constants $a_i, c_i$. The coefficient $a_i$ multiplies a 
non-normalizable metric deformation, and so is forced to zero by our choice 
of boundary conditions. The other integration constant $c_i$  multiplies 
an infinitesimal shift in the velocity of the brane. It turns out 
(see below) that a nonzero value for
$c_i$ leads to a nonzero value for $T_{0i}$ which violates our `renormalization' condition, consequently $c_i$ must be set to zero. In
summary,
\begin{equation}\label{metchangethreetwo}
\left(g^{(1)}_V\right)_{\alpha \beta} dx^\alpha dx^\beta= 2 \, r\, \partial_v \beta^{(0)}_i  \, dv \,dx^i.
\end{equation}

As in the scalar sector above, this solution grows by a factor of $r^3$ faster at the boundary than the shear zero mode. This slow fall off leads to
a divergent contribution to the stress tensor which precisely cancels an
equal and opposite divergence from  terms in the expansion of $g^{(0)}$ to first order in derivatives. As we will see below, the full contribution of $g^{(0)}+g^{(1)}$ to the vector part of the boundary stress tensor is just zero, again in agreement with our renormalization conditions.

\subsection{The symmetric tensors of $SO(3)$}
\label{ffive}

We now turn to $g^{(1)}_T$, the part of $g^{(1)}$ that transforms in the ${\bf 5}$, the symmetric traceless two tensor representation, of $SO(3)$. Let us parameterize our metric fluctuation by
\begin{equation}\label{symtraceeftwo}
\left(g^{(1)}_T\right)_{\alpha \beta} \,dx^\alpha \, dx^\beta = r^2 \,\alpha^{(1)}_{ij}(r)\, dx^i  \, dx^j,
\end{equation}
where $\alpha_{ij}$ is traceless and symmetric.
The Einstein's equation $E_{ij}=0$ yield
\begin{equation}\label{eomtenone}
{d\over dr}\left(r^5 \,f(r)\,{d\over dr} \alpha^{(1)}_{ij}
\right) = -{6 \, r^2}\, \sigma^{(0)}_{ij}\ ,
\end{equation}
where we have defined a symmetric traceless matrix
\begin{equation}
\sigma^{(0)}_{ij} = \partial_{(i } \beta^{(0)}_{j)}  -
{1 \over 3} \, \delta_{ij}\,\partial_m \beta^{(0)}_m \ .
\label{sijdef}
\end{equation}	

The LHS of \eqref{eomtenone} is the restriction of the 
operator $H$ of \eqref{homoop} to the tensor sector, and the RHS of 
this equation is the tensor part of the source term $s_1$. Note that 
$H$ is a first order operator in the variable 
$\alpha^{(1)'}_{ij}(r)$ and so may be integrated for an arbitrary source term.
The solution to this equation with arbitrary source term $s(r)$ is given by (dropping the tensor indices):
\begin{equation} \label{teneq} 
\alpha^{(1)}= -\int_r^\infty \frac{dx}{f(x) \, x^5} \int_1^x s(y) \, dy \ .
\end{equation}   
Note that the lower limit of the inner integral in \eqref{teneq} has been 
chosen to be unity. Provided that $s(x)$ is regular at $x=1$ (this is true 
of \eqref{eomtenone} and will be true at every order in perturbation theory),
$\int_1^x s(x)$ has a zero at $x=1$. It follows that the outer integrand 
in \eqref{teneq} is regular at nonzero $x$ (and in particular at $x=1$)
despite the explicit zero in the factor $f(x)$ in the denominator.
The solution for $\alpha^{(1)}$ is also normalizable provided the source 
is regular and grows at infinity slower than $r^3$. This condition is 
obeyed in \eqref{eomtenone} and is expected to continue to be obeyed 
at arbitrary order in perturbation theory (see the next section).

Applying \eqref{teneq} to the source term in \eqref{eomtenone} we find that 
the solution for $\alpha ^{(1)}_{ij}$ is given by 
\begin{equation}\label{metricperturbtwo}
(g^{(1)}_T)_{\alpha \beta} \,  dx^\alpha dx^\beta=  2\, r^2 \,  F(r)\,
\sigma^{(0)}_{ij}\,  dx^i dx^j.
\end{equation}
with 
\begin{equation}
\label{fdef}
F(r) = \int_r^{\infty}\, dx \,\frac{x^2+x+1}{x (x+1) \left(x^2+1\right)} ={1\over 4}\, \left[\ln\left(\frac{(1+r)^2(1+r^2)}{r^4}\right) - 2\,\arctan(r) +\pi\right] 
\end{equation}	
At large $r$ it evaluates to
\begin{equation}\label{metpertfintwo}
(g^{(1)}_T)_{\alpha \beta}\,  dx^\alpha dx^\beta
=  2\,  \left( r - {1 \over 4\, r^2} \right) \, 
\sigma^{(0)}_{ij} \, dx^i dx^j.
\end{equation}
As in the previous subsections, the first term in  \eqref{metpertfintwo} yields a contribution to the stress tensor that diverges like $r^3$, but  precisely cancels the corresponding divergence from first derivative terms in the expansion of $g^{(0)}$. However the second term in this expansion yields
an important finite contribution to the stress tensor, as we will see below.

\paragraph{Summary of the first order calculation:}
In summary, our final answer for $g^{(0)}+g^{(1)}$, expanded to first order
in boundary derivatives about $y^\mu=0$,  is given explicitly as
\begin{equation}\label{fullmettwo}
\begin{split}
ds^2 &=  2\, dv \, dr -r^2f(r) \, dv^2 + r^2 \, dx_i \, dx^i  \\
&- 2\, x^{\mu}\, \partial_{\mu}\beta^{(0)}_i \, dr\, dx^i  -
2\, x^\mu \, \partial_\mu \, 
\beta^{(0)}_ir^2(1-f(r)) \, dv \, dx^i   - 4 \frac{ x^\mu \partial_\mu \bz}{r^2} \, dv^2\\
& +2\, r^2\, F(r) \, \sigma^{(0)}_{ij} \, dx^i \, dx^j +{2\over 3}\, r \,\partial_i\beta^{(0)}_i \, dv^2 + 2\,  r \, \partial_v \beta^{(0)}_i \,dv \, dx^i .
\end{split}
\end{equation}

This metric solves  Einstein's equations to first order in the neighbourhood of $x^\mu=0$ provided the functions $\bz$ and $\beta^{(0)}_i$ satisfy
\begin{equation}\label{eomsatisfy}
\begin{split}
\partial_v \bz &=  {\partial_i \beta^{(0)}_i \over 3} \\
\partial_i \bz & =\partial_v \beta^{(0)}_i .
\end{split}
\end{equation}
%

\subsection{Global solution to first order in derivatives}
\label{fglobal}

In the previous subsection we have computed the metric $g^{(1)}$ about $x^\mu$
assuming that $\bz=1$ and $\beta^{(0)}_i=0$ at the origin. Since it is possible
to choose coordinates to set an arbitrary velocity to zero and an arbitrary
$\bz$ to unity at any given point (and since our perturbation procedure
is ultralocal), the results of the previous subsection contain enough information to write down the metric $g^{(1)}$ about any point. A simple way
to do this is to construct a covariant metric\footnote{By abuse of notation, we will 
refer to expressions transformation covariantly in the boundary metric (chosen here to be $\eta_{\mu\nu}$) as covariant. In particular, we are not interested in full bulk covariance as we will continue to restrict attention to a specific coordinatization of the fifth direction.}, as a function of $u_\mu$ and
$b$, which reduces to \eqref{fullmettwo} when $\bz=1$ and $\beta^{(0)}_i=0$.
It is easy to check that
\begin{equation}\label{replmet}
\begin{split}
ds^2 &= -2\, u_{\mu}\, dx^{\mu} dr -r^2\, f(b\,r)\, u_{\mu}u_{\nu}\, dx^{\mu}dx^{\nu} + r^2\, P_{\mu\nu}\, dx^{\mu} dx^{\nu} \\
&+2\, r^2 \,b\, F(b\, r)\, \sigma_{\mu\nu}  \, dx^{\mu} dx^{\nu} 
+{2\over 3} \, r \, u_{\mu}u_{\nu} \,\partial_{\lambda} u^{\lambda} \, dx^{\mu}dx^{\nu} -  r\, u^{\lambda}\partial_{\lambda}\left(u_\nu u_{\mu}\right)\, dx^{\mu} dx^{\nu} ,
\end{split}
\end{equation}
does the job, up to terms of second or higher order in derivatives. Here we have written the metric in terms of $\sigma_{\mu\nu}$ defined in \eqref{defst} and the function $F(r)$ introduced in \eqref{fdef}. Furthermore, it is easy to check that the metric above is the unique choice respecting the symmetries (again up to terms of second or higher order in derivatives). It follows that \eqref{replmet} is the metric $g^{(0)}+g^{(1)}$. It is also easily verified that the covariant version of \eqref{eomsatisfy} is \eqref{conszeroeth}. We will
interpret this as an equation of stress energy conservation in the next
subsection.

\subsection{Stress tensor to first order}
\label{fstress}
Given the solution to the first order equations, we can utilize the AdS/CFT dictionary to construct the boundary stress tensor using the prescription of \cite{Balasubramanian:1999re} (see also \cite{Henningson:1998gx}). For the metric \eqref{replmet} it  is not difficult to compute the stress tensor; all we need to do is compute the extrinsic curvature tensor $K_{\mu\nu}$ to the surface at fixed $r$. By convention, we choose the unit normal to this surface to be outward pointing, \ie\ pointing towards the boundary, in the definition of $K_{\mu\nu}$.  Using then the definition
\begin{equation}\label{formulaforst}
T^{\mu}_\nu = -2\lim_{r \to \infty}
r^4 \left( K^{\mu}_{\nu} -\delta^\mu_\nu \right) , 
\end{equation}
on our solution \eqref{replmet}, we find  the result is given simply as 
\begin{equation}\label{tmn}
T^{\mu \nu} ={1 \over b^4}
\left( 4\,  u^\mu u^\nu +\eta^{\mu \nu}
\right)  - {2 \, \over b^3} \, \sigma^{\mu \nu} .
\end{equation}
where $\sigma^{\mu\nu}$ was defined in \eqref{defst} and all
field theory indices are raised and lowered with the boundary metric
$\eta_{\mu \nu}$. As explained in the introduction, this stress tensor
implies that the ratio of viscosity to entropy density of our fluid is
$1/(4 \pi)$. Note that as mentioned previously, the expression \eqref{tmn} is only correct up to first derivative terms in the temperature ($T = 1/b$) and velocities.

\section{The metric and stress tensor at second order}
\label{metstrs}

In order to obtain the metric and stress tensor at second order
in the derivative expansion, we follow the method outlined in \sec{pertexp}
and implemented in detail in \sec{metstrf} to leading order. Concretely, we choose coordinates such that $\beta^{(0)}_i=0$ and $\bz=1$ at the point $x^\mu=0$. The metric $g^{(0)}+g^{(1)}$  given in \eqref{replmet} may be expanded to second order in derivatives. This involves Taylor expanding $g^{(0)}$ to second order and $g^{(1)}$ to first order, the second order analogue of  \eqref{fote}. As we have explained
in \sec{outsecond}, at this stage we also make the substitution $ \bz \rightarrow \bz+\bo$, and treat $\bo$ as an order $\varepsilon$
term, and so retain only those expressions that are of first derivative
order in $\bo$ (and contain no other derivatives). This process is straightforward and  we will not record the (rather lengthy) resultant expression here. To this expression we add the as yet undetermined metric fluctuation
\begin{equation} \label{metto}
\begin{split}
g^{(2)}_{\alpha \beta} \, dx^\alpha dx^\beta
&= -3 \, h_2(r) \, dv \, dr + r^2 \, h_2(r) \, dx_i \, dx^i + \frac{k_2(r)}{r^2}\,  dv^2 
 + 2 \, \frac{j^{(2)}_{i}(r)} {r^2}\,  dv \, dx^i + r^2 \alpha^{(2)}_{ij}\,  dx^i \, dx^j .\\
\end{split}
\end{equation}

We plug this metric into Einstein's equations and obtain a set of linear
second order differential equations that determine $h_2, k_2, j^{(2)}_i, \alpha^{(2)}_{ij}$. As in the previous section, $SO(3)$ symmetry ensures that
the equations for the scalars $h_2, k_2$, the vectors $j^{(2)}_i$, and the
tensor $\alpha^{(2)}_{ij}$ do not mix. Moreover, as we have explained in
\sec{pertexp}, the equations that determine these unknown functions are identical
to their first order counterparts in the homogeneous terms, but differ from
those equations in the sources. As a result, the only new calculation we
have to perform in order to obtain the metric at second order is the computation
of the source terms. Once these terms are available, the corresponding equations
may easily be integrated, as in the previous section.

Recall that the input metric into Einstein's equations includes terms that
arise out of the Taylor expansion of $g^{(0)}+g^{(1)}$ that have explicit factors
of the coordinates $x^\mu$. Nonetheless, a very simple argument assures
us that the source terms in the equations that determine $g^{(2)}$ must all
be independent of $x^\mu$. The argument runs as follows: We have explicitly
constructed $g^{(1)}$ in the previous section so that $E_{MN}\left(g^{(0)}+g^{(1)}\right) = O_{MN}$ where $O_{MN}$ is a local expression constructed out of second order or higher $x^\mu$ derivatives of velocity and temperature fields. It follows that $x^\mu$ dependence of sources, which may be obtained by Taylor expanding $O_{MN}$ about $x^\mu=0$, occurs
only at the three derivative level or higher. It follows that source terms
at the two derivative level have no $x^\mu$ dependence. Clearly, this
argument has a direct analogue at arbitrary order in perturbation theory.

A crucial input into the argument of the last paragraph was the fact that
$g^{(0)}+g^{(1)}$ satisfies Einstein's equations in a neighbourhood of $x^\mu=0$ (and not just at that point). As we have seen in the previous section, the
fact that the energy conservation equation is obeyed at $x^\mu=0$ allows
us to express all first derivatives of temperature in terms of first
derivatives of velocities (see \eqref{momcons} and \eqref{scaefcons}).
In addition, $\beta^{(0)}_i$ and $\bz$ must be chosen so that
\eqref{quad} is satisfied. The sixteen equations \eqref{quad}
can be grouped into sets that transform under $SO(3)$ as two scalars,
three vectors and one tensor (\ie, 5). We will now explain how these
constraints may be used to solve for 16 of the independent expressions
of second order in derivatives of velocity and temperature fields.

In order to do this, let us first list all two derivative `source' terms
that can be built out of second derivatives of $\bz$ or $\beta^{(0)}_i$, or
out of squares of first derivatives of $\beta^{(0)}_i$. These expressions may be separated
according to their transformation properties under $SO(3)$ as scalars,
vectors and tensors and higher order terms. The higher order pieces
will not be of interest to us. An exhaustive list of
these expressions that transform in the {\bf 1}, {\bf 3} or {\bf  5} is given in Table 1.\footnote{Note that the tensors are symmetric in their indices. The symmetrization as usual is indicated by parentheses. 
}
\begin{table}[h]\label{tablevtdefs}
\begin{center}
\begin{tabular}{|l|l|l|}
\hline
& & \\
\;\;\; {\bf 1} of $SO(3)$ 
&\;\;\;\;\; {\bf 3} of $SO(3)$ 
&\;\;\; \;\;\;\;\;{\bf 5} of $SO(3)$ \\
& &\\
\hline \hline
& &\\
\;\;\;$\Ss{1} = \frac{1}{b}\, \partial_v^2 b$ & 
\;\;\;$\V{1}_i = \frac{1}{b}\, \partial_i\partial_v b$ & 
\;\;\;$\T{1}_{ij} =\frac{1}{b}\,  \partial_i\partial_j b -\frac{1}{3} \, \Ss{3}\, \delta_{ij}$ \\
& &\\
\;\;\;$\Ss{2} = \partial_v\partial_i\beta_i$ &
\;\;\;$\V{2}_i = \partial_v^2\beta_i$ & 
\;\;\;$\T{2}_{ij} = \partial_{(i} \l_{j)}$ \\
& &\\
\;\;\;$\Ss{3} = \frac{1}{b}\, \partial^2 b$ &
\;\;\;$\V{3}_i = \partial_v \l_i $ & 
\;\;\;$\T{3}_{ij} = \partial_v \sigma_{ij}$ \\
& &\\
\;\;\;$\ST{1} = \partial_v\beta_i\,\partial_v\beta_i$ & 
\;\;\;$\V{4}_i = \frac{9}{5}\partial_j \sigma_{ji} - \partial^2\beta_i$ & 
\;\;\;$\TT{1}_{ij} = \partial_v\beta_i\,\partial_v\beta_j - \frac{1}{3}\, \ST{1}\, \delta_{ij}$ \\
& &\\
\;\;\;$\ST{2} = \l_i\,\partial_v\beta_i$ & 
\;\;\;$\V{5}_i = \partial^2\beta_i$ & 
\;\;\;$\TT{2}_{ij} = \l_{(i} \, \partial_v\beta_{j)} - \frac{1}{3}\,\ST{2}\, \delta_{ij} $ \\
& &\\
\;\;\;$\ST{3} =  \left(\partial_i\beta_i\right)^2$ & 
\;\;\;$\VT{1}_i = \frac{1}{3}(\partial_v\beta_i)(\partial_j\beta^j)$ &
\;\;\;$ \TT{3}_{ij} =2\, \epsilon_{kl(i} \, \partial_v\beta^k\, \partial_{j)} 
\beta^l+ \frac{2}{3} \,\ST{2}\, \delta_{ij}$ \\
& &\\
\;\;\;$\ST{4} = \l_i\,\l^i $ & 
\;\;\;$\VT{2}_i = -\epsilon_{ijk}\, \l^j\, \partial_v\beta^k $ & 
\;\;\;$\TT{4}_{ij} = \partial_k\beta^k\, \sigma_{ij} $ \\
& &\\
\;\;\;$\ST{5} = \sigma_{ij}\,\sigma^{ij}$ & 
\;\;\;$\VT{3}_i = \sigma_{ij}\,\partial_v\beta^j$ & 
\;\;\;$\TT{5}_{ij} = \l_i \,\l_j - \frac{1}{3}\, \ST{4}\, \delta_{ij}$ \\
& &\\
& 
\;\;\;$\VT{4}_i = \l_i\, \partial_j\beta^j$ & 
\;\;\;$\TT{6}_{ij} = \sigma_{ik}\, \sigma^k_j - \frac{1}{3}\, \ST{5}\, \delta_{ij}$ \\
\;\;\;& &\\
& 
\;\;\;$\VT{5}_i =  \sigma_{ij}\,\l^j$ & 
\;\;\;$ \TT{7}_{ij} = 2\, \epsilon_{mn(i} \, l^m \,\sigma^n_{j)} $\\
& &\\
\hline
\end{tabular} 
\caption{An exhaustive list of two derivative terms in made up from the temperature and velocity fields. In order to present the results economically, we have dropped the superscript on the velocities $\beta_i$ and the inverse temperature $b$, leaving it implicit that these expressions are only valid at second order in the derivative expansion.}
\end{center}
\end{table}
We define the vector $\l_i$ as the curl of the velocity \ie,
\begin{equation}
\l_i = \epsilon_{ijk}\, \partial^j\beta^k \ ,
\label{lsdefs}
\end{equation}	
and the symmetric traceless tensor $\sigma_{ij}$ has been previously defined in \eqref{sijdef}.

As a simple check on the completeness of expressions in Table 1,
 notice that the number of degrees of freedom in those of the tabulated 
expressions that are formed from a product of two single derivatives is 5 (in the
scalar sector), 5 $\times$ {\bf 3} (in the vector sector), and 7
$\times$ {\bf 5} in the tensor sector, leading to a total of 55 real parameters.  
Together with degrees of freedom from the two {\bf 7}s and one {\bf 9} that can also 
be formed from the product of two derivatives (but will play no role in our analysis) 
this gives 78 degrees of freedom. This is in agreement with the expected 
${1\over 2}\times 12 \times 13$ =78 ways of getting a symmetric object from twelve 
parameters (the first derivatives of the velocity fields). On the other hand, the 
genuinely two derivative terms in Table 1 have 
$3 \times {\bf 1} + 5 \times {\bf 3} +3 \times {\bf 5}=33$ degrees of freedom which 
together with a two derivative term that transforms in the {\bf 7} (which however plays 
no role in our analysis) is the expected number $40 =10 \times 4$ of  two derivative 
terms arising from temperature and velocity fields.

Assuming that we have already employed the first order conservation equation 
\eqref{conszeroeth} to eliminate the first derivatives of $b$, we have to deal with the constraint equation \eqref{quad} at the second order. Using the list of second order quantities given in Table 1, it is possible to show that \eqref{quad} take the form of the following linear relations between these two derivative terms:
\begin{equation}\label{constrainteq}
\begin{split}
\Ss{1} &= \frac{1}{3}\, \Ss{3} -\ST{1} + \frac{1}{9}\, \ST{3} +\frac{1}{6}\,\ST{4} -\frac{1}{3}\, \ST{5} \\
\Ss{2} &= \Ss{3} - \ST{1} +\frac{1}{2}\, \ST{4}- \ST{5}\\
\V1_i &= \frac{10}{9}\, \V4_i + \frac{1}{9}\, \V5_i + \frac{1}{3}\, \VT1_i  - \frac{1}{3} \, \VT2_i - \frac{2}{3}\, \VT3_i\\
\V2_i &= \frac{10}{9}\, \V4_i + \frac{1}{9} \, \V5_i -\frac{2}{3}\,\VT1_i + \frac{1}{6}\,\VT2_i - \frac{5}{3}\,\VT3_i\\
\V3_i &= -\frac{1}{3}\, \VT4_i + \VT5_i\\
\T{1}_{ij} &= \T{3}_{ij} +\TT{1}_{ij} + \frac{1}{3}\, \TT{4}_{ij} + \frac{1}{4}\,\TT{5}_{ij} + \TT{6}_{ij} \ .
\end{split}
\end{equation}
Given these relations we now proceed to analyze the potential source terms arising from the metric \eqref{replmet} at $\CO\left(\varepsilon^2\right)$. The analysis, as before, can be done sector by sector -- the computations for the scalar, vector and tensor sectors are given in \sec{sscalar}, \sec{svector} and \sec{sfive}, respectively.

\subsection{Solution in the scalar sector}
\label{sscalar}

Given the general second order fluctuation \eqref{metto}, we parameterize scalar components of $g^{(2)}$ in terms of the functions $h_2(r)$ and $k_2(r)$ according to
\begin{equation} \label{scalarefperttwo}
\begin{split}
 g^{(2)}_{ii} (r)& = 3 \, r^2\, h_2(r)\\
g^{(2)}_{vv}(r) &= \frac{k_2(r)}{r^2}\\
g^{(2)}_{vr}(r)&=-\frac{3}{2} h_2(r) \ .
\end{split}
\end{equation}
As we have explained in the \sec{fscalar}, the constraint Einstein's equations in this sector are given by the $r$ and $v$ component of the one-form formed by contracting the Einstein tensor with the vector dual to the one-form $dr$.  The $v$ component of this constraint, \ie\ the second order expansion of \eqref{scalarconstraint}, evaluates to
\begin{equation}\label{contwov}
\frac{1}{\bz} \partial_v \bo = \frac{1}{\bo}\,\ST{5} \ .
\end{equation}
This equation enables us to solve for the first $v$ derivative of $\bo$
in terms of two derivative terms made up of $\beta^{(0)}_i$, but imposes no further constraints on $\bz, \beta^{(0)}_i$. \eqref{contwov} has
a simple physical interpretation; it is simply the time component of the conservation  equation for the stress tensor \eqref{tmn}, expanded to second order in derivatives.
Consequently \eqref{contwov} is the Navier Stokes equation!

The $r$ component of the constraint, \ie\ \req{scaconstCtwo}, gives us one relation  between the functions $h_2(r)$ and $k_2(r)$ and their derivatives. As in \sec{fscalar}, to this constraint we must add one dynamical equation. We obtain the following equations
\begin{equation}\label{scalartwoeq}
\begin{split}
 5 \, h_2'(r) + r \, h_2''(r) &=  S_h(r)\\
 k_2'(r) &= S_k(r) \\
 &=   12 \,r^3 \,h_2(r) +(3 \,r^4-1) \,h_2'(r) + \widehat{S}_k(r)\ , 
\end{split}
\end{equation}
where
\begin{equation}\label{}
\begin{split}
S_h(r) & \equiv\frac{1}{3\, r^3}\, \ST{4} + \frac{1}{2}\, W_h(r) \, \ST{5}  \\
\widehat{S}_k(r)& \equiv -\frac{4 \,r}{3}\, \Ss{3} + 2 \,r \,\ST{1}  -\frac{2\, r}{9}\, \ST{3}+ \frac{1 + 2\, r^4}{6\, r^3} \,\ST{4} + \frac{1}{2}\,W_k(r) \,\ST{5}\ .
\end{split}
\end{equation}	
The functions $W_h(r)$ and $W_k(r)$ are given by
\begin{equation*}
\begin{split}
 W_h(r) &=\frac{4}{3}\; \frac{\left(r^2+r+1\right)^2-2 \left(3 \,r^2+2 r+1\right) \, F(r)   }{ r \,\left(r+1\right)^2\,\left(r^2+1\right)^2 } \ ,\\
 W_k(r) &=\frac{2}{3}\;\frac{4\, \left(r^2+r+1\right)\,\left(3\,r^4-1\right)\, F(r) - \left(2 r^5+2 r^4+2 r^3-r-1\right)}{ r\,\left(r+1\right)\,\left(r^2+1\right) } \ .
\end{split}
\end{equation*}
As advertised, it is clear that the differential operator acting on the functions $h_2(r)$ and $k_2(r)$ is identical to the one encountered in the first order computation in \sec{fscalar}. The equation \eqref{scalartwoeq} can be explicitly integrated; to do so it is useful to record the leading large $r$ behaviour of the source term $S_h(r)$:
\begin{equation}\label{asymscalarh}
S_h(r) \to {1\over r^3} \, S^{\infty}_h \equiv {1\over r^3} \,\left(\frac{1}{3}\, \ST{4} + \frac{2}{3}\,\ST{5} \right)\ .
\end{equation}
The first equation in \eqref{scalartwoeq} can be integrated given this asymptotic value to obtain the leading behaviour of the function $h_2(r)$. One finds
\begin{equation}\label{h2solution}
h_2(r) = - \frac{1}{4\, r^2}\, S^{\infty}_h+\int_r^\infty {dx \over x^5}\; \int_x^\infty dy\ y^4 \, \left(S_h(y) - {1\over y^3}\, S_h^{\infty} \right) . 
\end{equation}
The integral expression above can be shown to be of $\CO\left(r^{-5}\right)$ and hence the asymptotic behaviour of $h_2(r)$ is controlled by $s_h$.
Given $h_2(r)$, one can integrate up the second equation of \eqref{scalartwoeq} for $k_2(r)$. The leading large $r$ behaviour of the source term $S_k(r)$ is given by
\begin{equation}\label{asymscalark}
S_k(r) \to r\, S^{\infty}_k \equiv r\, \left(-\frac{4 }{3} \,\Ss{3} + 2  \,\ST{1}-\frac{2}{9}\, \ST{3} -\frac{1}{6}\, \ST{4}+\frac{7}{3} \,\ST{5}\right)  \ ,
\end{equation}
and hence we have
\begin{equation}\label{k2solution}
k_2(r) = \frac{r^2}{2}\, S^{\infty}_k-\int_r^\infty dx\, \left(S_k(x) - x\, S_k^{\infty}\right) \ .
\end{equation}
In this case the integral makes a subleading contribution starting at $\CO\left(r^{-1}\right)$. As in \sec{fscalar}, we have chosen the coefficients of homogeneous solutions to this differential equation so as to ensure normalizability and vanishing scalar contribution to the stress tensor (according to our
renormalization conditions).

\subsection{Solution in the vector sector}
\label{svector}

The analysis of the vector fluctuations at second order mimics the computation described in \sec{fvector}. The vector fluctuation in $g^{(2)}$ is chosen as described in \eqref{metto} to be 
\begin{equation}\label{vectorefperttwo}
g^{(2)}_{vi} = {j_i^{(2)} \over r^2} \ .
\end{equation}	
Once again, the analysis is easily done by looking at the constraint equations which are obtained by contracting the tensor $E_{MN}$ with the vector dual to $dr$. The 
$i^{{\rm th}}$ constraint equation evaluates to
\begin{equation}\label{vectorconstraint}
18\,\partial_i \bo= 5\,\V4_i + 5\, \V5_i + 15\, \VT1_i - \frac{15}{4}\,\VT2_i -\frac{33}{2}\,\VT3_i \ .
\end{equation}
This equation allows us to solve for the spatial derivatives of $\bo$ in
terms of derivatives of $\beta^{(0)}_i$ and $\bz$. \eqref{vectorconstraint}
is simply the expansion to second order in derivatives of the conservation of momentum of the stress tensor \eqref{tmn}.

To complete the solution in the vector channel, we need to solve for $j^{(2)}(r)$, which can be shown to satisfy a  dynamical equation
\begin{equation}\label{vectorequation}
\frac{d}{dr}\left({ 1\over r^3} \, {d\over dr} j^{(2)}_i(r)\right) = 
{\bf B}_i(r)  .
\end{equation}
Note that the LHS of this expression has the vector part of the operator $H$
acting on $j^{(2)}$. Here ${\bf B}_i(r)$ is the source term which is built out
of the second derivative terms transforming in the ${\bf 3}$ of $SO(3)$ 
given in 
Table 1.
\begin{equation} \label{bi}
{\bf B}(r) = \frac{p(r) \,  {\bf B}^{\infty} + {\bf B}^{{\rm fin}}}{18\, r^3\,  (r+1) \, \left(r^2+1\right)} 
\end{equation}
with
\begin{equation}\label{bidefs}
\begin{split}
{\bf B}^\infty  &= 4\, \left(10\, \V{4} +  \V{5} +3\, \VT{1} -3 \,\VT{2} -6\,\VT{3} \right)\\
{\bf B}^{{\rm fin}} & = 9\, \left(20\,  \V{4}- 5 \,\VT{2} - 6\,\VT{3}\right) ,
\end{split}
\end{equation}	
and we have introduced the polynomial:
\begin{equation}\label{pfndefs}
p(r) =2 \,r^3+2 r^2 +2\,r -3 \ .
\end{equation}	
Clearly $p(r)$ determines the large $r$ behaviour of the vector perturbation; asymptotically ${\bf B}(r) \to {1\over 9\, r^3} \, {\bf B}^\infty$.
Hence, integrating \eqref{vectorequation} we find that $j^{(2)}(r)$ is given as 
\begin{equation}\label{vectorsolution}
j^{(2)}_i(r) = -{r^2\over 36} \, {\bf B}^\infty_i + \int_r^\infty dx \; x^3\; \int_x^\infty dy \,\left({\bf B}_i(y) -  {1\over 9\, y^3} \,{\bf B}^\infty_i\right) \ ,
\end{equation}
where once again we have chosen the coefficients of homogeneous modes in
order to maintain normalizability and our renormalization condition. As with the first order computation described in \sec{fvector}, the solution \eqref{vectorsolution} makes no contribution to the stress tensor of the field theory.

\subsection{Solution in the tensor sector}
\label{sfive}

Finally, we turn to the tensor modes at second order where we shall recover the explicit form of the second order contributions to the stress tensor. Our task is now to determine the functions $\alpha^{(2)}_{ij}(r)$ in \eqref{metto}. As in \sec{ffive}, in the symmetric traceless sector of $SO(3)$ one has only the dynamical equation given by
\begin{equation}\label{tensorequation}
\frac{1}{2\, r}\frac{d}{d r}\left[r^5\left(1-\frac{1}{r^4}\right)\,\frac{d}{dr}\alpha_{ij}^{(2)}(r)\right] ={\bf A}_{ij}(r)
\end{equation}
where
\begin{equation*}
{\bf A}_{ij}(r) = {\bf a}_1(r)\, \left(\TT{1}_{ij}+ \frac{1}{3}\TT{4}_{ij} + \T{3}_{ij}\right) + {\bf a}_5(r) \,\TT{5}_{ij} + {\bf a}_6(r)\, \TT{6}_{ij}- \frac{1}{4}\, {\bf a}_7(r) \, \TT{7}_{ij} 
\end{equation*}
with the coefficient functions 
\begin{equation}\label{sourcecoef}
\begin{split}
{\bf a}_1(r) & = \frac{3\, p(r) +11}{p(r) +5} - 3\, r \, F(r) \\ 
{\bf a}_5(r) & =  \frac{1}{2} \left(1+\frac{1}{r^4}\right)\\
{\bf a}_6(r) & = \frac{4}{r^2}\;\frac{r^2 \, p(r) +3\, r^2 - r -1}{p(r) +5} - 6\, r\, F(r) \\
{\bf a}_7(r) &  =2\; \frac{p(r) +1}{p(r) +5} - 6\, r \, F(r)  \ .
\end{split}
\end{equation}
The functions $F(r)$ and $p(r)$ are defined in \eqref{fdef} and \eqref{pfndefs}, respectively. 

The desired solution to \eqref{tensorequation} can be found by intergrating the right hand side of the equation twice and choosing the solution to the homogenous solution such that we retain regularity\footnote{We have imposed the 
requirement that all metric functions are well behaved in its neighbourhood of 
$r=1$, a regular point in the spacetime manifold. Note that $r=1$ will not 
represent the horizon of our perturbed solution, but may well lie very near 
this horizon manifold.}  at $r =1$ and appropriate normalizability at infinity. The solution with these properties is 
\begin{equation*}
\alpha_{ij}^{(2)}(r) = -\int_r^\infty \frac{dx}{x \, (x^4 -1)}\; \int_1^x \,dy\; 2\,y \,{\bf A}_{ij}(y) 
\end{equation*}

The quantity of prime interest to us is the leading large $r$ behaviour of $\alpha^{(2)}_{ij}$. This can be inferred from the expressions for the coefficient functions given in \eqref{sourcecoef} and evaluates to
\begin{equation}\label{alphatwolarger}
\begin{split}
\alpha_{ij}^{(2)}(r) &= \frac{1}{r^2} \, \left(-\frac{1}{2} \,\TT{7}_{ij} + \TT{6}_{ij}  -\frac{1}{4}\, \TT{5}_{ij}\right)\\ 
&+  \frac{1}{2 r^4}\, \left[\left(1 - \frac{\ln 2}{2}\right)\,\left(\frac{1}{3}\,\TT{4}_{ij} + \TT{1}_{ij} + \T{3}_{ij}\right) +\frac{\ln2}{4}\, \TT{7}_{ij}+ \TT{6}_{ij}\right]
\end{split}
\end{equation}
The leading term here will give a divergent contribution to the stress tensor, which is necessary to cancel the divergence arising from the expansion of $g^{(0)} + g^{(1)}$ to second order. The subleading piece in \eqref{alphatwolarger} is the term that will provide us with the second order stress tensor. Before proceeding to evaluate the stress tensor we  present the full solution to second order, appropriately covariantized.

\subsection{Global solution to second order in derivatives}
\label{sglobal}
Consider the following metric 
\begin{equation}\label{replmettnd}
ds^2 = -2\, u_{\mu}\, dx^{\mu} dr -r^2\, f(b\,r)\, u_{\mu}u_{\nu}\, dx^{\mu}dx^{\nu} + r^2\, P_{\mu\nu}\, dx^{\mu} dx^{\nu} + 3\, b^2 \,h_2(b\, r) \,u_{\mu}\,dx^\mu dr+\CG_{\mu \nu} \, dx^\mu dx^\nu  \ ,
\end{equation}
where we have defined a symmetric tensor $\CG_{\mu \nu}$ by combing the contributions in the field theory directions from the first and second order metrics $g^{(0)} + g^{(1)}$ 
\begin{equation}\label{replG}
\begin{split}
\CG_{\mu \nu} &=r^2 \,\left( 2 \,b \, F(b\, r)\,\sigma_{\mu\nu} 
+b^2 \,\alpha^{(2)}_{\mu \nu}(b \,r) \right) + \frac{1}{r^2} \,\left(  \frac{2}{3}\,  r^3\, \partial_{\lambda} u^{\lambda}\, u_{\mu}\, u_{\nu}+\frac{k_2(b\, r)}{b^2}\, u_\mu \, u_\nu
\right)  \\
& \qquad + r^2\, b^2 \, h_2(b\, r)\, P_{\mu\nu} +\frac{1}{r^2} \,\left( - 2\, r^3 \, \CD u_{\alpha} + \frac{1}{b^2}\, j^{(2)}_\alpha(b\, r)\right)  \,P^\alpha_\nu\, u_{\mu} \ .
\end{split}
\end{equation}
The covariant expression for $\alpha^{(2)}_{\mu\nu}$ is given by \eqref{tensorequation} with the replacements
\begin{equation}\label{tensordef}
\begin{split}
\TT{1}_{ij} & \;\; \rightarrow \;\; \left(T_{2d}\right)_{\mu \nu}\ , \qquad 
\TT{4}_{ij} \;\; \rightarrow \;\;\left(T_{2c} \right)_{\mu \nu}\ , \qquad 
\TT{5}_{ij} \;\; \rightarrow \;\; \l_\mu \,\l_\nu - \frac{1}{3} P_{\mu \nu} \,\l^{\alpha} \l_{\alpha}\\
\TT{6}_{ij}& \;\; \rightarrow \;\; \left(T_{2b}\right)_{\mu \nu}\ , \qquad 
\TT{7}_{ij}  \;\; \rightarrow \;\; 2\, \left(T_{2a}\right)_{\mu \nu}\ , \qquad
\T{3}_{ij}  \;\; \rightarrow \;\; \left(T_{2e}\right)_{\mu \nu}\ .
\end{split}
\end{equation}
Further, $j^{(2)}_\mu$ given by \eqref{vectorsolution} with ${\bf B}_i(r)
\rightarrow B_\nu(r)$, where ${\bf B}_i(r)$ is given by
\eqref{bi} and we make the following replacements 
\begin{equation} \label{vecrep}
\begin{split}
& \V{4}_i \;\; \rightarrow \;\;\frac{9}{5}\left[P^\alpha_\nu \, P^{\beta\gamma}\,\partial_\gamma\, \partial_{(\beta} u_{\alpha)}  - \frac{1}{3} \,P^{\alpha\beta}\, P^\gamma_{\;\nu}\, \partial_\gamma  \partial_\alpha u_\beta \right] -  P^\mu_ \nu\, P^{\alpha\beta}\, \partial_\alpha\partial_\beta u_{\mu}\\
& \V{5}_i \;\; \rightarrow \;\; P^\mu_ \nu\, P^{\alpha\beta}\partial_\alpha\partial_\beta u_{\mu}\\
&\VT{1}_i \;\;\rightarrow\;\; \partial_\alpha u^\alpha \, \CD u_\nu \ , \qquad 
\VT{2}_i \;\;\rightarrow\;\; \epsilon_{\alpha\beta\gamma\nu} \, u^\alpha\, \CD u^\beta \,\l^\gamma\ , \qquad 
\VT{3}_i \;\;\rightarrow\;\; \sigma_{\alpha\nu}\, \CD u^\alpha\ .
\end{split}
\end{equation}
Finally, $h_2(r)$ and $k_2(r)$ are given by \eqref{h2solution} and \eqref{k2solution} respectively,  and in the functions  $S_h(r)$, $S_k(r)$,  $S_h^\infty$ and $S_k^\infty$ defined in \eqref{asymscalarh} and \eqref{asymscalark} we are required to make  the replacements
\begin{equation} \label{scalrep}
\begin{split}
& \Ss{3}  \;\;\rightarrow\;\; \frac{1}{\bz} 
P^{\alpha \beta}\,\partial_\alpha \partial_\beta \bz\, \qquad 
\ST{1}  \;\;\rightarrow\;\; \CD u^\alpha \, \CD u_\alpha \ , \qquad 
\ST{2} \;\;\rightarrow\;\;  \l_\mu \, \CD u^\mu \\
&\ST{3} \;\;\rightarrow\;\; (\partial_\mu u^\mu)^2\ , \qquad 
\ST{4} \;\;\rightarrow\;\; \l_\mu \,\l^\mu\ , \qquad
\ST{5} \;\;\rightarrow\;\; \sigma_{\mu\nu}\,\sigma^{\mu\nu} \ .
\end{split}
\end{equation}

It may be checked that this metric is the
unique (up to terms that differ at third or higher order in derivatives) 
covariant expression that reduces to two derivative
solution determined in the previous subsections, in the neighbourhood
of any point $y^\mu$ after making the coordinate change that
sets $\bz=1$ and $\beta^{(0)}_i=0$ at that point. It follows that
\eqref{replmettnd} is the desired metric $g^{(0)}+g^{(1)}+g^{(2)}$.

\subsection{Stress tensor to second order}
\label{sstress}
The stress tensor dual to the solution to second order described in \sec{sglobal} can be obtained by using the standard formula \eqref{formulaforst}. To determine the extrinsic curvature at large $r$, it suffices to know the asymptotic form of the metric since we are interested in terms that have a finite limit as we take $r \to \infty$. Consequently, in order to compute the stress tensor it is sufficient to replace the various functions of $r$ that have appeared in the computation in \sec{sscalar}, \sec{svector} and \sec{sfive} by their large $r$ asymptotics. The stress tensor may the be computed in a straightforward fashion, yielding
\begin{equation}\label{sttwoder}
\begin{split}
\left(T_2\right)_{vv}  &=\left(T_2\right)_{vi} = 0 \ , \\
\left(T_2\right)_{ij} &= -\frac{\ln2}{4} \; \TT{7}_{ij} - \TT{6}_{ij} + \left(-1 + \frac{\ln \, 2}{2}\right)\, \left(\T{3}_{ij} + \TT{1}_{ij} + \frac{1}{3} \,\TT{4}_{ij}\right) \ .
\end{split}
\end{equation}
The vanishing of $\left(T_2\right)_{v\mu}$ is actually guaranteed by our renormalization condition. It is easy to check that the covariant form of the expression \eqref{sttwoder} is indeed the stress tensor quoted in \eqref{fmst}.
This result is the main prediction of our fluctuation analysis to second order in the derivative expansion.

\section{Second order fluid dynamics}
\label{secfluid}
In the previous section we have derived the precise form of the fluid
dynamical stress tensor dual to gravity on \AdS{5} including all terms
with no more than two derivatives. In this section we initiate a study of
the physics of this stress tensor. In \sec{secweyl} below we will demonstrate
that our stress tensor transforms homogeneously under Weyl transformations.
In \sec{smallfluc} we compute the dispersion relation for low frequency
sound and shear waves that follows from our stress tensor. In \sec{sechydro} we compare our predictions for the transport coefficients at second order with those of 
\cite{Baier:2007fj}.

\subsection{Weyl transformation of the stress tensor}
\label{secweyl}

Thus far we have extracted the stress tensor for a conformal fluid in flat space $\R^{3,1}$. We would like to ensure that the second order stress tensor we have derived transforms homogeneously under Weyl rescaling. In order to check this 
we perform the obvious minimal covariantization of our stress tensor to 
generalize it to a fluid stress tensor about an arbitrary boundary metric  
 $g_{\mu \nu}$.\footnote{All metrics in this subsection refer to the metric on the boundary, \ie., the background spacetime on which the fluid is propagating.}
and study its Weyl transformation properties. 
 
Consider the Weyl transformation of the boundary metric
\begin{equation} \label{conftransf} \begin{split}
 g_{\mu\nu} &= e^{2\phi}\, \widetilde{g}_{\mu\nu} \;\; \Rightarrow \;\; 
  g^{\mu\nu} = e^{-2\phi}\widetilde{g}^{\mu\nu} \\
\& \;\;\; u^\mu& = e^{-\phi}\, \widetilde{u}^\mu \ , \qquad 
T= e^{-\phi} \widetilde{T}  .
\end{split}
\end{equation}

It is well known that the first order truncation of the stress tensor
\eqref{fmst} transforms as $T^{\mu \nu} = e^{-6 \,\phi}\, {\widetilde T}^{\mu \nu}$
under this transformation (see for instance Appendix D of \cite{Wald:1984gr}). We proceed to show  that this transformation rule holds for the two
derivative stress tensor as well. This transformation property, together with the tracelessness of the stress tensor,  ensures Weyl invariance of the fluid dynamical equations $\nabla_\mu T^{\mu \nu}$, appropriate for a conformal fluid.

It follows from \eqref{conftransf} that $P^{\mu\nu}=g^{\mu\nu}+u^{\mu}u^{\nu}=e^{-2\phi}\, \widetilde{P}^{\mu\nu}$.
The Christoffel symbols transform as \cite{Wald:1984gr}
\begin{equation*}
 \Gamma_{\lambda\mu}^{\nu}= \widetilde{\Gamma}_{\lambda\mu}^{\nu} + \delta^{\nu}_{\lambda}\, \partial_{\mu}\phi+ \delta^{\nu}_{\mu}\, \partial_{\lambda}\phi- \widetilde{g}_{\lambda\mu}\,\widetilde{g}^{\nu\sigma}\, \partial_{\sigma}\phi\,.
\end{equation*}
The transformation of the covariant derivative of $u^{\mu}$ is given by
\begin{equation}
 \nabla_{\mu}u^{\nu}=\partial_{\mu}u^{\nu}+\Gamma_{\mu\lambda}^{\nu}\, u^{\lambda}
  =e^{-\phi}\, \left[\widetilde{\nabla}_{\mu}\, \widetilde{u}^{\nu}+\delta^{\nu}_{\mu}\, \widetilde{u}^{\sigma}\, \partial_{\sigma}\phi- \widetilde{g}_{\mu\lambda}\, \widetilde{u}^{\lambda}\, \widetilde{g}^{\nu\sigma}\, \partial_{\sigma}\phi\right].
\end{equation}
This equation can be used to derive the transformation of various quantities of interest in fluid dynamics, such as the acceleration $a^\mu$, shear $\sigma^{\mu \nu}$, \etc. 
\begin{equation}
\begin{split}
\theta &= \nabla_{\mu}u^{\mu}
  =e^{-\phi}\, \left({\widetilde\nabla}_{\mu} {\widetilde u}^{\mu}+3\, \widetilde{u}^{\sigma}\, \partial_{\sigma}\phi\right) = e^{-\phi} \, \left(\widetilde{\theta} + 3\, \widetilde{\CD} \phi \right) \ ,\\
 a^{\nu}&= \CD u^\nu = u^{\mu}\nabla_{\mu}u^{\nu}
  =e^{-2\phi}\left(\widetilde{a}^{\nu}+\widetilde{P}^{\nu\sigma}\, \partial_{\sigma}\phi\right),\\
 \sigma^{\mu\nu} &= P^{\lambda(\mu} \nabla_\lambda u^{\nu)}   - \frac{1}{3} \, P^{\mu\nu}\,\nabla_{\lambda}u^{\lambda} 
= e^{-3\phi} \; \widetilde{\sigma}^{\mu\nu},\\
\l^{\mu} &=  u_{\alpha}\,\epsilon^{\alpha \beta \gamma \mu}
\nabla_\beta u_{\gamma} = e^{-2 \phi}\, {\widetilde \l^\mu}
 \end{split}
 \label{fderdefs}
\end{equation}
where in the last equation we have accounted for the fact that all epsilon symbols
in \eqref{defst} should be generalized in curved space to their covariant
counterparts. The objects with correct tensor transformation properties scale as metric determinants \ie,    $\epsilon_{\alpha \beta \gamma \delta}  \propto \sqrt{g}$, and  $\epsilon^{\alpha \beta \gamma \delta}\propto \frac{1}{\sqrt{g}}$, from which it is easy to infer their scaling behaviour under conformal transformations; in particular, $\epsilon_{\alpha \beta\gamma \delta} = e^{4\phi} \; {\widetilde \epsilon}_{\alpha \beta \gamma \delta}$ and  $\epsilon^{\alpha \beta\gamma \delta} = e^{-4\phi} \;  {\widetilde \epsilon}^{\,\alpha \beta \gamma \delta}$. 

The Weyl transformation of the two derivative terms that occur in the stress tensor \eqref{defst} is given by
\begin{equation}
\begin{split}
T_{A}^{\;\mu\nu} &= e^{-4\phi} \; \widetilde{T}_{A}^{\;\mu\nu}\ , \qquad \qquad \qquad  \;\;\;\, {\rm for \;} A = \{2a, 2b\} \\
T_{B}^{\;\mu\nu} &=  e^{-4\phi}\; \left({\widetilde T}_{B}^{\;\mu\nu} +\widetilde {\delta T}_{B}^{\;\mu\nu}\right)\ , \qquad {\rm for \;} B = \{2c, 2d, 2e\} 
\end{split}
\end{equation}
where the inhomogeneous terms arising in the Weyl transformation are:
\begin{equation} \label{inhomo}
\begin{split}
\delta T_{2c}^{\;\mu\nu} &= 3 \,  \CD \phi \, \left( \nabla^{(\mu}  u^{\nu)} + u^{(\mu} \,a^{\nu)} - \frac{1}{3} \, \theta \, P^{\mu \nu} \right)\\
\delta T_{2d}^{\;\mu\nu} &= 2\, a^{(\mu} \,\nabla^{\nu)} \phi + 2\, u^{(\mu}\,a^{\nu)} \, \CD \phi -\frac{2}{3} \,a^\alpha\,\nabla_\alpha \phi \\ 
& \qquad+ 2\, u^{(\mu}\, \nabla^{\nu)}\phi \, \CD \phi + u^{\mu}\, u^{\nu} \, \left(\CD \phi \right)^2 -\frac{1}{3}\,P^{\mu \nu} \,  \left(\CD \phi \right)^2 + \nabla^\mu \phi\, \nabla^\nu \phi - \frac{1}{3} \, P^{\mu \nu} \, \nabla^{\alpha} \phi \, \nabla_\alpha \phi\\
\delta T_{2e}^{\;\mu\nu} &= -\nabla^{(\mu}u^{\nu)} \, \CD \phi - 3\, u^{(\mu} a^{\nu)} \, \CD \phi + \frac{1}{3}\, P^{\mu \nu}\, \theta\, \CD \phi - 2\, a^{(\mu} \,\nabla^{\nu)}\phi + \frac{2}{3}\,P^{\mu \nu} \, a^\alpha \, \nabla_\alpha \phi \\
& \qquad - u^\mu \,u^\nu (\CD \phi)^2 + \frac{1}{3}\, P^{\mu \nu}\, (\CD \phi)^2 - 2\, u^{(\mu} \, \nabla^{\nu)} \phi \, \CD \phi - \nabla^\mu \phi\, \nabla^\nu \phi  + \frac{1}{3} \, P^{\mu \nu} \, \nabla_\alpha \phi \, \nabla^\alpha \phi
\end{split}
\end{equation}

While the conformal transformation involves the inhomogeneous terms presented in \eqref{inhomo} we need to ensure that the full stress tensor is Weyl covariant. Satisfyingly, these inhomogeneous terms cancel among themselves in
the precise combination that  occurs in \eqref{fmst}; consequently the linear
combination of terms that occurs in the stress tensor transforms covariantly.
Note that the cancelation of inhomogeneous terms depends sensitively on the
ratio of coefficients of $T_{2c}$, $T_{2d}$ and $T_{2e}$; and so provides a check of our results. Note however that $T_{2a}$ and $T_{2b}$ are separately Weyl covariant. In summary, our result for the two derivative stress tensor is a linear combination (with precisely determined coefficients) of three independently Weyl covariant forms, with scaling weight $-4$ (for upper indices).

Using the transformation of the temperature \eqref{conftransf}
it follow that the full stress tensor transforms under Weyl transformation as
\begin{equation}
T^{\mu \nu} = e^{-6\phi} \, \widetilde{T}^{\mu \nu} \ . 
\label{strweylcov}
\end{equation}	
R. Loganayagam \cite{loga} informs us that he has found a compact way of rewriting our  stress tensor $T^{\mu \nu}$ \eqref{fmst} that makes the Weyl invariance of each of its three pieces  manifest. 

\subsection{Spectrum of small fluctuations}
\label{smallfluc}
Consider a static bath of homogeneous fluid at temperature $T$. Given the 
two derivative stress tensor derived above \eqref{defst}, it is trivial to solve for the spectrum of small oscillations of fluid dynamical modes about this background. As the background is translationally invariant, these fluctuations can be taken to have the form
\begin{equation}\label{linfluc}
\begin{split}
\beta_i(v,x^j) &= \delta \beta_i\, e^{i \,\omega \,v  + i \, k_j  x^j}\\
T(v,x^j) &= 1 + \delta T \, e^{i\,\omega \, v+ i\, k_j x^j}\\
\end{split}
\end{equation}
Plugging \eqref{linfluc} into the equations of fluid dynamics \eqref{encon}, and working
to first order in $\delta \beta_i$ and $\delta T$, these equations reduce
to a set of four homogeneous linear equations in the amplitudes
$\delta \beta_i$ and $\delta T$. The coefficients of these equations
are functions of $\omega$ and $k_i$. These equations have nontrivial solutions
if and only if the matrix formed out of these coefficient functions has
zero determinant. Setting the determinant of the matrix of coefficients to
zero one can find the following two dispersion relations:
\begin{equation}\label{sound}
{\rm Sound \;mode:} \qquad {\boldsymbol{\omega}} ({\bf k}) = \pm\frac{{\bf k} }{\sqrt{3}} +\frac{i\,
   {\bf k} ^2}{6}\pm\frac{ (3-\ln 4)}{24 \sqrt{3}}\,{\bf k} ^3 +
{\cal O}\left({\bf k}^4\right) , 
\end{equation}

\begin{equation}\label{shear}
{\rm Shear \;mode:} \qquad {\boldsymbol{\omega}} ({\bf k} ) = \frac{i \,{\bf k} ^2}{4}+\frac{i}{32}\, \left(2-\ln 2\right)  \, {\bf k}^4
+ {\cal O}\left({\bf k}^6\right) ,
\end{equation}

where we have defined the rescaled energy and momenta
\begin{equation} \label{bold}
{\boldsymbol{\omega}} = \frac{\omega}{\pi\, T}\ , \qquad {\bf k}= \frac{k}{\pi \, T }\  .
\end{equation}

It would be interesting to check our prediction against the quasinormal 
mode analysis of \cite{Kovtun:2005ev}.

\subsection{Comparison with Baier et.al. (added in v2)}
\label{sechydro}

In this subsection we compare our results with those of the 
preprint \cite{Baier:2007fj} which appeared in an arXiv listing simultaneously
with the first version of this paper. Where our results overlap we 
find perfect agreement. 

The authors of \cite{Baier:2007fj} demonstrated that conformal invariance determines the two derivative fluid dynamical stress tensor of any conformal field th    eory to upto five parameters (see also \cite{loga} for an alternate 
derivation of this result). The five undetermined parameters are  the 
coefficients of the various Weyl invariant two derivative expressions built out
of velocity. These coefficients which are named $\tau_\Pi$, $\kappa$ and 
$\lambda_{1,2,3}$ by \cite{Baier:2007fj}, are defined via the equation\footnote{Note that  $\sigma^{\mu\nu}$ as defined in \eqref{defst} differs from that of \cite{Baier:2007fj} by a factor of $2$ and we define $\omega^{\mu \nu} = -\Omega^{\mu\nu}$. We have introduced the tilded quantities $\tilde{\tau}_\Pi$ \etc, to account for the normalization difference. We have $\tilde{\tau}_\Pi = 2\, \tau_\Pi$, $\tilde{\lambda}_1 = 4 \, \lambda_1$ and $\tilde{\lambda}_2 = -2\,\lambda_2$.}
\begin{equation}
T_{(2)}^{\mu \nu} = \tilde{\tau}_\Pi \,\eta\, \CT_1^{\mu \nu} + \kappa \, \CT_2^{\mu \nu} + \tilde{\lambda}_1 \, \CT_3^{\mu \nu}  + \tilde{\lambda}_2 \, \CT_4^{\mu \nu} + \lambda_3 \, \CT_5^{\mu \nu} \ , 
 \label{bsecost}
\end{equation}	
with\footnote{We have used the notation of \cite{Baier:2007fj} to indicate the symmetric, transverse tracelessness; for any two tensor $\CF^{\mu \nu}$ 
\begin{equation}
\CF^{\langle \mu \nu\rangle} = P^{\mu \alpha}\,P^{\nu \beta}\, \CF_{(\alpha \beta)} - \frac{1}{3}\, P^{\mu \nu}\,P^{\alpha\beta}\, \CF_{\alpha\beta} \nonumber
\label{}
\end{equation}	
}
\begin{equation} \label{relation}
\begin{split}
\CT_1^{\mu \nu}  &= ^{\langle}\!\CD \sigma^{\mu \nu \rangle} + \frac{1}{3}\, \theta\, \sigma^{\mu \nu} \equiv\frac{1}{3} \,T_{2c}^{\mu \nu} +T_{2d}^{\mu \nu}+T_{2e}^{\mu \nu} \ , \\
\CT_2^{\mu \nu} &=R^{\langle\mu\nu \rangle}  - 2  \, R^{\alpha \langle\mu \nu \rangle\beta}\,  u_\alpha \, u_\beta \ , \\
\CT_3^{\mu \nu} & = \sigma^{\lambda \langle\mu} \, \sigma^{\nu \rangle}_{\;\;\lambda} \equiv  T^{\mu \nu}_{2b} \ , \\
\CT_4^{\mu \nu} & = \sigma^{\lambda \langle\mu} \, \omega^{\nu \rangle}_{\;\;\lambda}\equiv \frac{1}{2}\, T_{2a}^{\mu\nu} \\
\CT_5^{\mu \nu} & = \omega^{\lambda \langle\mu} \, \omega^{\nu \rangle}_{\;\;\lambda} \ . 
\end{split}
\end{equation}	
Here the quantities $\theta$ and $\sigma$ are defined previously 
\eqref{fderdefs} and $\omega^{\mu \nu} = -P^{\mu \alpha}\, P^{\nu \beta}\, 
\partial_{[\alpha}u_{\beta]}$ is the antisymmetric two tensor built from 
velocity derivatives. In \eqref{relation} we have also reexpressed the 
operators $\CT_1^{\mu \nu}, \cdots , \CT_5^{\mu \nu}$ of \cite{Baier:2007fj} as linear combinations 
of the tensors we have used in our paper.  These relations are easy to verify 
given the definitions \eqref{defst} and \eqref{fderdefs}.

The analysis of \cite{Baier:2007fj} was able to determine three of the five coefficients above. Specifically, for ${\cal N}=4$ Yang Mills they find:
\begin{equation}
\tilde{\tau}_\Pi = \frac{2 - \ln 2}{\pi \,T} \ , \qquad \kappa = \frac{\eta}{\pi \, T} \ , \qquad \tilde{\lambda}_1 = \frac{2\,\eta}{\pi\,T}  \ , \qquad \eta = \frac{\pi}{8}\,N^2 \, T^3
\label{betalres}
\end{equation}	
The coefficients $\lambda_2$ and $\lambda_3$ were undetermined by their 
analysis.
Translating the results of our paper for the second order stress tensor 
into the language of this subsection and reinstating the overall normalization\footnote{Recall that in writing \eqref{defst} we scaled out an overall factor of $16\,\pi\, G_N$, which evaluates to $\frac{N^2}{8\pi^2}$ for $\CN =4$ SYM. The value for any other conformal field theory with gravitational dual is simply determined by the central charge.} we find
\begin{equation}
\tilde{\tau}_\Pi = \frac{2 - \ln 2}{\pi\, T} \ , \qquad \tilde{\lambda}_1 = \frac{2\,\eta}{\pi \,T} \ , \qquad \tilde{\lambda}_2 = \frac{2\,\eta\, \ln 2}{\pi\,T}\ , \qquad \lambda_3=0.
\label{}
\end{equation}	
As the coeffient $\kappa$ does not enter the equations of fluid dynamics in 
flat space, our analysis leaves this coefficient undetermined. 

In summary our results for $\tau_\Pi$ and $\lambda_1$ are in agreement 
with those of \cite{Baier:2007fj}. Putting together the results of 
our paper with the value of $\kappa$ determined in \cite{Baier:2007fj}
we have a prediction for the full two fluid dynamical stress tensor of a 
conformal fluid dual to gravity, 
propagating on an arbitrary curved background. 

We can also directly compare the dispersion relations presented in the 
previous subsection with those obtained in \cite{Baier:2007fj}. 
Our sound wave dispersion relation \eqref{sound}  agrees with the direct 
computation of quasinormal modes presented in  \cite{Baier:2007fj}. 
The shear mode dispersion relation of the previous subsection also 
agrees with the direct quasinormal mode analysis of  \cite{Baier:2007fj}
upto the order that it should, \ie. upto terms of order $k^3$. 
Note that the $k^4$ contribution to the  dispersion of the shear 
quasinormal mode of \cite{Baier:2007fj}  does not 
agree with the coefficient of $k^4$ in \eqref{shear}, but (as explained 
in \cite{Baier:2007fj}) there is no reason that it should. As terms of order 
$k^4$ are two orders subleading compared to terms of order $k^2$ (the order 
at which the first order stress tensor contributes to the shear quazinnormal 
mode) the coefficient of $k^4$ in the shear quasinormal mode potentially 
receives contributions from third order terms in the fluid dynamics stress 
tensor that we have not accounted for in \eqref{shear}.

\section{Discussion}
\label{discuss}

We have demonstrated how to start from a general, stationary black
 brane solution describing perfect fluid dynamics and promote the parameters 
in the gravitational solution to physical fluctuation modes. This procedure 
allows us to set up a fluctuation analysis which can be used to extract 
the boundary stress tensor of fluids dual to gravity in asymptotically 
\AdS{5} spacetimes, in a derivative expansion. Our procedure is ultralocal: 
we obtain our solution by patching together local tubes of the black brane 
solution into a global solution of Einstein's equations. The fact that 
our solutions tubewise approximate black branes (see  \cite{Janik:2006gp, 
Janik:2006ft,Heller:2007qt} for related observations) is the 
gravitational analogue of the fact that the fluid  dynamics approximation only works when the fluid is in local equilibrium.  We find this  structure of our solutions quite fascinating and  feel that it might have the potential to teach us important lessons about black brane dynamics. 

Equation \eqref{defst} is a prediction for the stress tensor of all four dimensional conformal fluids that admit a dual gravitational description. As we have described 
in the introduction, there exists an infinite number of examples of conformal 
field theories with a gravitational dual that differ substantially in their 
field content, spectrum of  operators,\ etc. Nonetheless, up to an overall normalization, each  of these theories has the same fluid dynamical expansion! Consequently, the fluid dynamics described in this paper has a degree of universality 
associated with it. At the one derivative level, the fluid stress tensor 
has a single undetermined parameter - the shear viscosity. The value of 
$\eta$ that we find is in agreement with earlier work, $\eta/s = 1/(4\,\pi)$. 
This relationship has been shown to have a larger degree of universality 
than is apparent from our work; it applies  to all field theories, whether conformal or not, that have a gravitational  dual.  This relationship has also been conjectured to act as a lower bound  on the viscosity of a relativistic field theory. It would be interesting 
to investigate whether any of the new two derivative coefficients we have 
found in this paper display extended universality features and also whether they are sensitive to higher derivative terms as discussed recently for the shear viscosity to entropy ratio in \cite{Brigante:2007nu,Kats:2007mq}.

As we have remarked in \sec{preamble}, it would be interesting to 
investigate whether our result for the stress tensor is consistent with 
the so called Israel-Stewart formulation of fluid dynamics \cite{loga}, 
a framework that has been employed in several practical investigations 
of fluid flows. 

Relatedly, we note that recent claims \cite{Romatschke:2007mq} that the RHIC plasma violate the viscosity to entropy bound referred to above are based on the analysis of RHIC plasma flows using first order fluid dynamics. However, a satisfactory analysis of these flows should include contributions from higher order 
terms in the fluid dynamical expansion. It is possible that the stress tensor 
derived in this paper will be useful in this regard.\footnote{We thank O. Aharony
for this suggestion.}

It may be possible to use the formalism presented in this paper to obtain 
a better understanding of the formal structure of the fluid dynamical 
expansion of quantum field theories. In this context it is useful to 
recall that the spectrum of regular small oscillations about a uniform 
black brane hosts an infinite spectrum of quasinormal modes. In this paper 
we have effectively constructed the `chiral Lagrangian' corresponding to 
those of the quasinormal fluctuations that are Goldstone modes (and so have 
zero frequency when at zero $k$). The remaining quasinormal modes played no 
role in our analysis, as they are nonperturbatively massive in the inverse 
temperature ($\omega \sim T = 1/b$). The existence of these non perturbative 
modes probably implies that the fluid dynamical expansion is asymptotic 
rather than convergent, and might allow us to predict the location of the 
first singularity in the Borel transform of this perturbation series. 

Recall that metric fluctuation, for any asymptotically \AdS{5} solution to Einstein's equations, decays at large $r$ like $1/r^4$ relative to the background. The coefficients of this $1/r^4$ decay are functions of the four field theory coordinates $x^\mu$; in a particular gauge these functions may be identified with the 9 components of the traceless boundary stress tensor. This stress tensor is constrained to obey the equations of energy momentum conservation, but is otherwise unconstrained by local analysis. The Fefferman-Graham \cite{Fefferman:1985ci} method (or equivalently the formalism of holographic renormalization, see 
\cite{Skenderis:2002wp,Papadimitriou:2004ap} for reviews)
demonstrates that any such conserved stress
tensor, regarded as a boundary condition to Einstein's equations, leads
to a unique and well defined power series expansion (in $1/r$) of
an asymptotically AdS metric. Local analysis near the boundary thus 
appears to indicate that the space of solutions to Einstein's equations in 
AdS space is parameterized by the set of all conserved energy momentum 
tensors in four dimensions. This would be  very surprising from the dual 
field theory viewpoint, as a set of four equations does not define a 
well posed initial value problem for nine functions.

The results of our paper suggest a (perhaps not unanticipated) resolution
to this puzzle. In the derivative expansion in which we work,
all except a four function set of this naive nine function class of metrics
are unacceptably singular and so do not constitute a legal solution to 
Einstein's equations. Generic data result in singularities that develop 
at a finite value of $r$ ($r=1/b$ in our set up) and so are not easily visible 
in the Fefferman-Graham expansion, which is guaranteed to work only in   
an open neighbourhood of the boundary. The class of boundary stress tensors 
that generate acceptable metrics are parameterized by four functions 
($\beta_i(x^\mu)$ and $b(x^\mu)$) rather than nine. These four functions 
are further constrained to obey the four equations of stress energy 
conservation. As four equations constitute a well 
defined\footnote{Note that our notion of a well posed PDE system is 
simply that we do not have an under-constrained system of equations. We are 
not making any claim regarding the well posedness of generic initial data; 
only initial data in  the regime of our perturbation analysis together 
with the boundary conditions is guaranteed to lead to regular solutions. 
The general question of global regularity of Navier-Stokes equation is 
of course an interesting open problem.}  
initial value problem for a set of four functions, the set of legal
solutions to Einstein's equations are parameterized by data that consists 
of functions of 3 spatial rather than 4 spacetime boundary variables, in 
 agreement with field theory expectations. It would of course be of very great interest to understand how these results of the previous paragraph  generalize beyond the boundary derivative expansion. 

In this context it is also relevant to note that the equations of fluid
dynamics themselves develop singularities under certain situations.
It would be interesting to investigate the gravitational dual of this
process of singularity formation.\footnote{We thank D. Berenstein for
discussions on this question.}  More generally, the map from solutions 
of fluid dynamics to solutions of gravity could allow one to use the insight gained from the hundred year long study of the equations of fluid dynamics to understand 
qualitatively new gravitational solutions. For example, one might hope to learn about stationary inhomogeneous brane solutions (analogous to those discovered in the study of Gregory-Laflamme instabilities of black strings and branes) using the fluid dynamical description. 

As we have described, in this paper we have derived explicit 
formulae for the metric dual to any solution of the Navier-Stokes equations. 
We have not yet investigated the global structure of the resulting spacetime. 
It seems very plausible that (under suitable physical conditions) the 
spacetimes we have constructed have regular event horizons. The event horizon 
is a null surface; we  expect  it to closely approximate\footnote{We 
would like to thank H. Reall for many useful discussions on this point.}  
the surface $r\, b(x^\mu)=1$.  If this is the case we should be able to 
compute an explicit expression for this surface order by order in perturbation 
theory. It may then be possible to use our understanding of the horizon 
to define a locally positive divergence entropy current (a `pullback' of the one-form dual to the natural null generators of the horizon back to the boundary might play a role in such a construction).
In the most optimistic scenario such an exercise could relate classic results 
about the positivity of null congruence expansions (resulting from the 
Raychaudhuri's equation with the usual proviso of energy conditions) to 
the local positivity of entropy production in fluid dynamics; a result that 
would be of obvious interest. The language of dynamical horizons\footnote{Note that the homogeneous spacetime background being simply the uniform black brane, has a trapped surface; under the fluctuations we generically expect the 
trapped surface to be generated earlier in the radial evolution \ie,  
at a larger radial coordinate (assuming of course appropriate 
energy conditions).} \cite{Ashtekar:2003hk} may well prove to be the 
appropriate framework for such a discussion. We hope to return to 
this intriguing issue in the future.

Recall that the construction presented in this paper yields the gravitational 
dual of every solution of the equations of fluid dynamics. Standard field 
theory lore asserts that generic field theory evolutions are well described 
by solutions to the  equations of fluid dynamics in the regime of interest 
to this paper.  Consequently the AdS/CFT correspondence seems to imply that 
the construction described here yields the generic legal solution of gravity in \AdS{5}, within  its domain of applicability. If our guess of the previous paragraph is 
correct -- \ie, if all our solutions possess a regular event horizon that 
shield the boundary from the singularity -- then our results appear to be of 
relevance to the cosmic censorship conjecture 
\cite{Penrose:1969pc,Wald:1997wa}. However we emphasize that our analysis 
 applies only in a long wavelength expansion, and so presumably does not 
apply to several classes of scenarios that putatively violate this conjecture.
More physically, fluid dynamics applies only under the assumption of local 
thermal equilibrium. Presumably naked singularities (if they exist) 
are dual to `far from local equilibrium' boundary physics.

Several other natural generalizations of the work presented in this paper
immediately suggest themselves. First, the results of our paper are likely
to have an analogue for $d$ dimensional gravitational theories with
a negative cosmological constant for every $d\geq 4$. Second, it should
be possible to extend the results of our paper to spaces that asymptote
to \AdS{d+1} in global coordinates (and whose dual description is, therefore,
fluid dynamics on $S^{d-1} \times \R_t$). More ambitiously, it may be possible
to extend the results of our work to field theories whose spacetime
metric asymptotes to
$$ds^2={dr^2\over r^2} + r^2 \,ds_{bdy}^2$$
for a more general class of metrics $ds_{bdy}^2$. In particular, the
generalization to time dependent metrics would permit the study of the
gravitational dual of forced fluids, a subject of interest to the study
of turbulence. Finally, it should not be difficult to generalize our study
to two derivative theories of gravity interacting with gauge fields.
We expect that the dual description of such a system will be the fluid
dynamics of a system with a number of additional conserved charges (equal
to the number of commuting vector fields). Note however that unlike the
uncharged system, this charged fluid dynamics will
not be universal at nonlinear order, as gauged supergravities do not
in general admit a consistent truncation to the Einstein-Maxwell sector.
For instance couplings of the form $f(\phi) F_{\mu \nu} F^{\mu \nu}$, for
an arbitrary scalar field $\phi$, constitute a source for $\phi$; this is an 
effect that plays an important role in studies of the attractor mechanism.

Despite this non universality, IIB supergravity on \AdS{5}  $\times S^5$
(for instance) should be dual to a completely well defined charged fluid
dynamics. It would be of interest to use the methods of this
paper determine the form of this fluid dynamical stress tensor. Among other
things, this exercise would allow us  to zero in on the origin of the worrying
apparent discrepancy between the formulas of charged black hole thermodynamics
and the formulas of fluid dynamics, as reported in \cite{Bhattacharyya:2007vs}.

It should prove relatively straightforward to generalize the study of this 
paper to the fluid dynamics of non conformal backgrounds of gravity. For 
example, Scherk-Schwarz compactifications of AdS spaces yield a particularly
simple set of gravitational backgrounds dual to confining gauge theories. 
Indeed, a moment's thought is enough to convince oneself that the deconfined 
phase fluid dynamical stress tensor of the 2+1 dimensional confining  
gauge theory (dual to Scherk-Schwarz compactified ${\cal N}=4$ Yang Mills)  
is simply the dimensional reduction of the stress tensor of $d=4$, ${\cal N}=4$ 
Yang Mills (\ie, the stress tensor presented in this paper, \eqref{fmst}) plus a constant additive piece. This seemingly 
trival additive piece is physically very important; it leads to qualitatively 
new phenomena. For  instance, \cite{Aharony:2005bm,Lahiri:2007ae} have 
plasmaballs and plasmarings; static finite lumps of fluid with a boundary. 
Such configurations have qualitatively new classes of excitations; 
localized collective coordinates associated with fluctuations of the 
boundary. These new collective coordinates will interact
with those studied in this paper. If it proves to be 
technically possible, it would be fascinating to formulate and study the 
resulting dynamics.

\subsection*{Acknowledgements}

It is a pleasure to thank R. Loganayagam,  H. Reall and T. Wiseman for
collaboration at the initial stages of this project, and for several invaluable comments and
discussions throughout its execution. We thank the authors of 
\cite{Baier:2007fj} for sharing their results with us prior to publication. 
We would like to thank P.~Basu, D.~Berenstein, R.~Gopakumar,  S.~Gupta, 
M.~Headrick, H.~Liu, J.~Lucietti, G.~Mandal, S.~Trivedi, S.~Wadia  and all 
the students in the TIFR theory room for useful discussions. We would 
also like to thank O. Aharony, R. Gopakumar, S. Lahiri, N. Seiberg, A. Strominger, M. Van Raamsdonk, S. Wadia and T. Wiseman for comments on an advance version of this manuscript. We would also like to thank M. Headrick for his excellent Mathematica package for performing gravity computations. VH, SM and MR would like to thank the Issac Newton Institute, Cambridge for hospitality during the workshop, ``Strong Fields, Strings and Integrability" where this project was initiated. VH  and MR are supported in part by STFC. The work of SM was supported in part by a Swarnajayanti
Fellowship. Two of us (SB and SM) must also acknowledge our debt to the steady and
generous support of the people of India for research in basic science.

\bibliographystyle{utphys}

\begin{thebibliography}{10}

\bibitem{Policastro:2001yc}
G.~Policastro, D.~T. Son, and A.~O. Starinets, ``The shear viscosity of
  strongly coupled N = 4 supersymmetric Yang-Mills plasma,'' {\em Phys. Rev.
  Lett.} {\bf 87} (2001) 081601,
\href{http://www.arXiv.org/abs/hep-th/0104066}{{\tt hep-th/0104066}}.

\bibitem{Janik:2005zt}
R.~A. Janik and R.~Peschanski, ``Asymptotic perfect fluid dynamics as a
  consequence of AdS/CFT,'' {\em Phys. Rev.} {\bf D73} (2006) 045013,
\href{http://www.arXiv.org/abs/hep-th/0512162}{{\tt hep-th/0512162}}.

\bibitem{Janik:2006gp}
R.~A. Janik and R.~Peschanski, ``Gauge / gravity duality and thermalization of
  a boost- invariant perfect fluid,'' {\em Phys. Rev.} {\bf D74} (2006) 046007,
\href{http://www.arXiv.org/abs/hep-th/0606149}{{\tt hep-th/0606149}}.

\bibitem{Nakamura:2006ih}
S.~Nakamura and S.-J. Sin, ``A holographic dual of hydrodynamics,'' {\em JHEP}
  {\bf 09} (2006) 020,
\href{http://www.arXiv.org/abs/hep-th/0607123}{{\tt hep-th/0607123}}.

\bibitem{Bhattacharyya:2007vs}
S.~Bhattacharyya, S.~Lahiri, R.~Loganayagam, and S.~Minwalla, ``Large rotating
  AdS black holes from fluid mechanics,''
\href{http://www.arXiv.org/abs/0708.1770}{{\tt arXiv:0708.1770
[hep-th]}}.

\bibitem{Sin:2006pv}
S.-J. Sin, S.~Nakamura, and S.~P. Kim, ``Elliptic flow, Kasner universe and
  holographic dual of RHIC fireball,'' {\em JHEP} {\bf 12} (2006) 075,
\href{http://www.arXiv.org/abs/hep-th/0610113}{{\tt hep-th/0610113}}.

\bibitem{Janik:2006ft}
R.~A. Janik, ``Viscous plasma evolution from gravity using AdS/CFT,'' {\em
  Phys. Rev. Lett.} {\bf 98} (2007) 022302,
\href{http://www.arXiv.org/abs/hep-th/0610144}{{\tt hep-th/0610144}}.

\bibitem{Friess:2006kw}
J.~J. Friess, S.~S. Gubser, G.~Michalogiorgakis, and S.~S. Pufu, ``Expanding
  plasmas and quasinormal modes of anti-de Sitter black holes,'' {\em JHEP}
  {\bf 04} (2007) 080,
\href{http://www.arXiv.org/abs/hep-th/0611005}{{\tt hep-th/0611005}}.

\bibitem{Kajantie:2006ya}
K.~Kajantie and T.~Tahkokallio, ``Spherically expanding matter in AdS/CFT,''
  {\em Phys. Rev.} {\bf D75} (2007) 066003,
\href{http://www.arXiv.org/abs/hep-th/0612226}{{\tt hep-th/0612226}}.

\bibitem{Heller:2007qt}
M.~P. Heller and R.~A. Janik, ``Viscous hydrodynamics relaxation time from
  AdS/CFT,'' {\em Phys. Rev.} {\bf D76} (2007) 025027,
\href{http://www.arXiv.org/abs/hep-th/0703243}{{\tt hep-th/0703243}}.

\bibitem{Kajantie:2007bn}
K.~Kajantie, J.~Louko, and T.~Tahkokallio, ``Gravity dual of 1+1 dimensional
  Bjorken expansion,'' {\em Phys. Rev.} {\bf D76} (2007) 106006,
\href{http://www.arXiv.org/abs/0705.1791}{{\tt arXiv:0705.1791 [hep-th]}}.

\bibitem{Chesler:2007sv}
P.~M. Chesler and L.~G. Yaffe, ``The stress-energy tensor of a quark moving
  through a strongly-coupled N=4 supersymmetric Yang-Mills plasma: comparing
  hydrodynamics and AdS/CFT,''
\href{http://www.arXiv.org/abs/0712.0050}{{\tt arXiv:0712.0050
 [hep-th]}}.

\bibitem{Benincasa:2007tp}
P.~Benincasa, A.~Buchel, M.~P. Heller, and R.~A. Janik, ``On the supergravity
  description of boost invariant conformal plasma at strong coupling,''
\href{http://www.arXiv.org/abs/0712.2025}{{\tt arXiv:0712.2025
 [hep-th]}}.

\bibitem{Herzog:2002fn}
C.~P. Herzog, ``The hydrodynamics of M-theory,'' {\em JHEP} {\bf 12} (2002)
  026,
\href{http://www.arXiv.org/abs/hep-th/0210126}{{\tt hep-th/0210126}}.

\bibitem{Policastro:2002tn}
G.~Policastro, D.~T. Son, and A.~O. Starinets, ``From AdS/CFT correspondence to
  hydrodynamics. II: Sound waves,'' {\em JHEP} {\bf 12} (2002) 054,
\href{http://www.arXiv.org/abs/hep-th/0210220}{{\tt hep-th/0210220}}.

\bibitem{Policastro:2002se}
G.~Policastro, D.~T. Son, and A.~O. Starinets, ``From AdS/CFT correspondence to
  hydrodynamics,'' {\em JHEP} {\bf 09} (2002) 043,
\href{http://www.arXiv.org/abs/hep-th/0205052}{{\tt hep-th/0205052}}.

\bibitem{Son:2002sd}
D.~T. Son and A.~O. Starinets, ``Minkowski-space correlators in AdS/CFT
  correspondence: Recipe and applications,'' {\em JHEP} {\bf 09} (2002) 042,
\href{http://www.arXiv.org/abs/hep-th/0205051}{{\tt hep-th/0205051}}.

\bibitem{Herzog:2002pc}
C.~P. Herzog and D.~T. Son, ``Schwinger-Keldysh propagators from AdS/CFT
  correspondence,'' {\em JHEP} {\bf 03} (2003) 046,
\href{http://www.arXiv.org/abs/hep-th/0212072}{{\tt hep-th/0212072}}.

\bibitem{Herzog:2003ke}
C.~P. Herzog, ``The sound of M-theory,'' {\em Phys. Rev.} {\bf D68} (2003)
  024013,
\href{http://www.arXiv.org/abs/hep-th/0302086}{{\tt hep-th/0302086}}.

\bibitem{Kovtun:2003wp}
P.~Kovtun, D.~T. Son, and A.~O. Starinets, ``Holography and hydrodynamics:
  Diffusion on stretched horizons,'' {\em JHEP} {\bf 10} (2003) 064,
\href{http://www.arXiv.org/abs/hep-th/0309213}{{\tt hep-th/0309213}}.

\bibitem{Buchel:2003tz}
A.~Buchel and J.~T. Liu, ``Universality of the shear viscosity in
  supergravity,'' {\em Phys. Rev. Lett.} {\bf 93} (2004) 090602,
\href{http://www.arXiv.org/abs/hep-th/0311175}{{\tt hep-th/0311175}}.

\bibitem{Buchel:2004di}
A.~Buchel, J.~T. Liu, and A.~O. Starinets, ``Coupling constant dependence of
  the shear viscosity in N=4 supersymmetric Yang-Mills theory,'' {\em Nucl.
  Phys.} {\bf B707} (2005) 56--68,
\href{http://www.arXiv.org/abs/hep-th/0406264}{{\tt hep-th/0406264}}.

\bibitem{Buchel:2004qq}
A.~Buchel, ``On universality of stress-energy tensor correlation functions in
  supergravity,'' {\em Phys. Lett.} {\bf B609} (2005) 392--401,
\href{http://www.arXiv.org/abs/hep-th/0408095}{{\tt hep-th/0408095}}.

\bibitem{Kovtun:2004de}
P.~Kovtun, D.~T. Son, and A.~O. Starinets, ``Viscosity in strongly interacting
  quantum field theories from black hole physics,'' {\em Phys. Rev. Lett.} {\bf
  94} (2005) 111601,
\href{http://www.arXiv.org/abs/hep-th/0405231}{{\tt hep-th/0405231}}.

\bibitem{Kovtun:2005ev}
P.~K. Kovtun and A.~O. Starinets, ``Quasinormal modes and holography,'' {\em
  Phys. Rev.} {\bf D72} (2005) 086009,
\href{http://www.arXiv.org/abs/hep-th/0506184}{{\tt hep-th/0506184}}.

\bibitem{Benincasa:2005iv}
P.~Benincasa, A.~Buchel, and A.~O. Starinets, ``Sound waves in strongly coupled
  non-conformal gauge theory plasma,'' {\em Nucl. Phys.} {\bf B733} (2006)
  160--187,
\href{http://www.arXiv.org/abs/hep-th/0507026}{{\tt hep-th/0507026}}.

\bibitem{Maeda:2006by}
K.~Maeda, M.~Natsuume, and T.~Okamura, ``Viscosity of gauge theory plasma with
  a chemical potential from AdS/CFT,'' {\em Phys. Rev.} {\bf D73} (2006)
  066013,
\href{http://www.arXiv.org/abs/hep-th/0602010}{{\tt hep-th/0602010}}.

\bibitem{Mas:2006dy}
J.~Mas, ``Shear viscosity from R-charged AdS black holes,'' {\em JHEP} {\bf 03}
  (2006) 016,
\href{http://www.arXiv.org/abs/hep-th/0601144}{{\tt hep-th/0601144}}.

\bibitem{Saremi:2006ep}
O.~Saremi, ``The viscosity bound conjecture and hydrodynamics of M2-brane
  theory at finite chemical potential,'' {\em JHEP} {\bf 10} (2006) 083,
\href{http://www.arXiv.org/abs/hep-th/0601159}{{\tt hep-th/0601159}}.

\bibitem{Son:2006em}
D.~T. Son and A.~O. Starinets, ``Hydrodynamics of R-charged black holes,'' {\em
  JHEP} {\bf 03} (2006) 052,
\href{http://www.arXiv.org/abs/hep-th/0601157}{{\tt hep-th/0601157}}.

\bibitem{Benincasa:2006fu}
P.~Benincasa, A.~Buchel, and R.~Naryshkin, ``The shear viscosity of gauge
  theory plasma with chemical potentials,'' {\em Phys. Lett.} {\bf B645} (2007)
  309--313,
\href{http://www.arXiv.org/abs/hep-th/0610145}{{\tt hep-th/0610145}}.

\bibitem{Son:2007vk}
D.~T. Son and A.~O. Starinets, ``Viscosity, Black Holes, and Quantum Field
  Theory,''
\href{http://www.arXiv.org/abs/0704.0240}{{\tt arXiv:0704.0240 [hep-th]}}.

\bibitem{Shuryak:2003xe}
E.~Shuryak, ``Why does the quark gluon plasma at RHIC behave as a nearly ideal
  fluid?,'' {\em Prog. Part. Nucl. Phys.} {\bf 53} (2004) 273--303,
\href{http://www.arXiv.org/abs/hep-ph/0312227}{{\tt hep-ph/0312227}}.

\bibitem{Shuryak:2004cy}
E.~V. Shuryak, ``What RHIC experiments and theory tell us about properties of
  quark-gluon plasma?,'' {\em Nucl. Phys.} {\bf A750} (2005) 64--83,
\href{http://www.arXiv.org/abs/hep-ph/0405066}{{\tt hep-ph/0405066}}.

\bibitem{Shuryak:2006se}
E.~V. Shuryak, ``Strongly coupled quark-gluon plasma: The status report,''
\href{http://www.arXiv.org/abs/hep-ph/0608177}{{\tt hep-ph/0608177}}.

\bibitem{Gervais:1974db}
J.-L. Gervais and B.~Sakita, ``Quantized relativistic string as a strong
  coupling limit of the Higgs model,'' {\em Nucl. Phys.} {\bf B91} (1975)
301.

\bibitem{Baier:2007fj}
R.~Baier, P.~Romatschke, D.~T. Son, A.~O. Starinets, and M.~A. Stephanov,
  ``Relativistic viscous hydrodynamics, conformal invariance, and holography,''
\href{http://www.arXiv.org/abs/0712.2451}{{\tt arXiv:0712.2451 [hep-th]}}.

\bibitem{Muronga:2003ta}
A.~Muronga, ``Causal Theories of Dissipative Relativistic Fluid Dynamics for
  Nuclear Collisions,'' {\em Phys. Rev.} {\bf C69} (2004) 034903,
\href{http://www.arXiv.org/abs/nucl-th/0309055}{{\tt nucl-th/0309055}}.

\bibitem{Israel:1979wp}
W.~Israel and J.~M. Stewart, ``Transient relativistic thermodynamics and
  kinetic theory,'' {\em Ann. Phys.} {\bf 118} (1979)
341--372.

\bibitem{Andersson:2007gf}
N.~Andersson and G.~L. Comer, ``Relativistic Fluid Dynamics: Physics for Many
  Different Scales,'' {\em Living Reviews in Relativity} {\bf 10} (2007),
  no.~1,.

\bibitem{loga}
R.~Loganayagam, {\em Work in progress, to appear}.

\bibitem{Balasubramanian:1999re}
V.~Balasubramanian and P.~Kraus, ``A stress tensor for anti-de Sitter
  gravity,'' {\em Commun. Math. Phys.} {\bf 208} (1999) 413--428,
\href{http://www.arXiv.org/abs/hep-th/9902121}{{\tt hep-th/9902121}}.

\bibitem{Henningson:1998gx}
M.~Henningson and K.~Skenderis, ``The holographic Weyl anomaly,'' {\em JHEP}
  {\bf 07} (1998) 023,
\href{http://www.arXiv.org/abs/hep-th/9806087}{{\tt hep-th/9806087}}.

\bibitem{Wald:1984gr}
R.~Wald, ``General relativity,'' {\em Chicago, University of Chicago Press,
  1984, 504 p.} (1984).

\bibitem{Brigante:2007nu}
M.~Brigante, H.~Liu, R.~C. Myers, S.~Shenker, and S.~Yaida, ``Viscosity Bound
  Violation in Higher Derivative Gravity,''
\href{http://www.arXiv.org/abs/0712.0805}{{\tt arXiv:0712.0805 [hep-th]}}.

\bibitem{Kats:2007mq}
Y.~Kats and P.~Petrov, ``Effect of curvature squared corrections in AdS on the
  viscosity of the dual gauge theory,''
\href{http://www.arXiv.org/abs/0712.0743}{{\tt arXiv:0712.0743 [hep-th]}}.

\bibitem{Romatschke:2007mq}
P.~Romatschke and U.~Romatschke, ``Viscosity Information from Relativistic
  Nuclear Collisions: How Perfect is the Fluid Observed at RHIC?,'' {\em Phys.
  Rev. Lett.} {\bf 99} (2007) 172301,
\href{http://www.arXiv.org/abs/0706.1522}{{\tt arXiv:0706.1522 [nucl-th]}}.

\bibitem{Fefferman:1985ci}
C.~Fefferman and C.~Graham, ``Conformal invariants,'' {\em Elie Cartan et les
  Mathematiques d'Aujourd'hui, Asterisque} {\bf 95} (1985).

\bibitem{Skenderis:2002wp}
K.~Skenderis, ``Lecture notes on holographic renormalization,'' {\em Class.
  Quant. Grav.} {\bf 19} (2002) 5849--5876,
\href{http://www.arXiv.org/abs/hep-th/0209067}{{\tt hep-th/0209067}}.

\bibitem{Papadimitriou:2004ap}
I.~Papadimitriou and K.~Skenderis, ``AdS / CFT correspondence and geometry,''
\href{http://www.arXiv.org/abs/hep-th/0404176}{{\tt hep-th/0404176}}.

\bibitem{Ashtekar:2003hk}
A.~Ashtekar and B.~Krishnan, ``Dynamical horizons and their properties,'' {\em
  Phys. Rev.} {\bf D68} (2003) 104030,
\href{http://www.arXiv.org/abs/gr-qc/0308033}{{\tt gr-qc/0308033}}.

\bibitem{Penrose:1969pc}
R.~Penrose, ``Gravitational collapse: The role of general relativity,'' {\em
  Riv. Nuovo Cim.} {\bf 1} (1969)
252--276.

\bibitem{Wald:1997wa}
R.~M. Wald, ``Gravitational collapse and cosmic censorship,''
\href{http://www.arXiv.org/abs/gr-qc/9710068}{{\tt gr-qc/9710068}}.

\bibitem{Aharony:2005bm}
O.~Aharony, S.~Minwalla, and T.~Wiseman, ``Plasma-balls in large N gauge
  theories and localized black holes,'' {\em Class. Quant. Grav.} {\bf 23}
  (2006) 2171--2210,
\href{http://www.arXiv.org/abs/hep-th/0507219}{{\tt hep-th/0507219}}.

\bibitem{Lahiri:2007ae}
S.~Lahiri and S.~Minwalla, ``Plasmarings as dual black rings,''
\href{http://www.arXiv.org/abs/0705.3404}{{\tt arXiv:0705.3404 [hep-th]}}.

\end{thebibliography}

\providecommand{\href}[2]{#2}\begingroup\raggedright\endgroup

\end{document}